\documentclass[a4paper]{ar-1col}
\usepackage{arydshln}
\usepackage[numbers]{natbib}
\usepackage{bm}
\usepackage{url}
\usepackage{amsmath,amssymb,amsthm}
\usepackage{pifont}
\usepackage{subfigure}
\usepackage{multirow}
\usepackage{bbold}
\usepackage{simplewick}
\usepackage{enumitem}

\newcommand{\bra}[1]{\< #1 \right|}
\newcommand{\ket}[1]{\left| #1 \>}
\definecolor{purple}{rgb}{0.5, 0, 0.5}
\renewcommand{\>}{\right\rangle}
\newcommand{\<}{\left\langle}   \newcommand{\bit}{\begin{itemize}}
\newcommand{\eit}{\end{itemize}}    
\newcommand{\f}{\frac}
\newcommand{\be}{\begin{equation}}
\newcommand{\ee}{\end{equation}}
\newcommand{\ba}{\begin{align}}

\newcommand{\lf}{\left(}
\newcommand{\ri}{\right)}
\newcommand{\dd}{\mathrm{d}}

\newcommand{\Tr}{\operatorname{Tr}}

\newcommand{\seq}{s_\mathrm{eq}}

\newcommand{\kket}[1]{\left|\hspace{-0.28mm}\left| #1 \>\hspace{-0.55mm}\>}
    
\usepackage{xr}
\usepackage{amsmath}
\usepackage[ruled,vlined]{algorithm2e}
\usepackage[]{graphicx}
\usepackage{grffile}
\usepackage{mathrsfs}
\usepackage{framed}

 \usepackage{bbm}
 \usepackage{bm}
 \usepackage{braket}
 \usepackage{mathtools}

\definecolor{purple}{rgb}{0.5, 0.5, 0}

\usepackage[]{qcircuit}

\renewcommand{\vec}{\mathbf}

\jname{Submitted to Annual Review of Condensed Matter Physics}
\jyear{2022}

\begin{document}

\markboth{Fisher Khemani Nahum Vijay}{Random Quantum Circuits}

\title{Random Quantum Circuits}

\author{Matthew P. A. Fisher$^1$, Vedika Khemani$^2$, \\ Adam Nahum$^3$, Sagar Vijay$^1$
\affil{$^1$Department of Physics, University of California, Santa Barbara, CA 93106, USA}
\affil{$^2$Department of Physics, Stanford University, Stanford, CA 94305, USA}
\affil{$^3$Laboratoire de Physique, \'Ecole Normale Sup\'erieure, CNRS, Universit\'e PSL, Sorbonne Universit\'e, Universit\'e de Paris, 75005 Paris, France.}}

\begin{abstract}
Quantum circuits --- built from local unitary gates and local measurements --- are a new playground for quantum many-body physics and a tractable setting to explore universal collective phenomena far-from-equilibrium. These models have shed light on longstanding questions about thermalization and chaos, and on the underlying universal dynamics of quantum information and entanglement. In addition, such models generate new sets of questions and give rise to  phenomena with no traditional analog, such as new dynamical phases in quantum systems that are monitored 
by an external observer. Quantum circuit dynamics is also topical in view of experimental progress in building digital quantum simulators that allow control of precisely these ingredients. Randomness in the circuit elements allows a high level of theoretical control, with a key theme being mappings between real-time quantum dynamics and effective classical lattice models or dynamical processes. Many of the universal phenomena that can be identified in this tractable setting apply to much wider classes of more structured many-body dynamics.
\end{abstract}

\maketitle

\tableofcontents

\section{Introduction}
\label{sec:intro}

Quantum dynamical phenomena, such as the transport of conserved quantities, correlation and response functions of local observables, or dynamics of low-lying excitations, have traditionally been used to probe the universal properties of  quantum condensed matter at low temperatures and near equilibrium.
In contrast, since quantum coherence is
easily destroyed at high temperatures, 
it might seem that quantum matter evolved far from  its ground-state
would  fail to exhibit universal dynamics that is distinctively quantum. However, efforts to understand
out-of-equilibrium dynamics in
lattice models  \cite{DAlessio-Rigol2016_review, nandkishore2015many,abanin2019colloquium,serbyn2021quantum,moudgalya2021quantum,doyon2020lecture,Bulchandani_integrability_2021,fagotti2008evolution,buyskikh2016entanglement,lauchli2008spreading,kim2013ballistic,ho2017entanglement,bardarson2012unbounded,serbyn2013universal,huse2014phenomenology,nahum2017quantum,potter2021entanglement}
quantum field theories \cite{calabrese2005evolution,calabrese2009entanglement,asplund2015entanglement,liu2014entanglement}, and black holes \cite{hayden2007black,roberts2015localized,abajo2010holographic,hartman2013time,hubeny2007covariant}  have shown
that universal structures can emerge in  
quantum correlations and in the patterns of quantum entanglement of a many-body system.  These structures underlie thermalization, when it occurs, as well as novel forms of non-thermalizing dynamics \cite{DAlessio-Rigol2016_review, nandkishore2015many}.

The search for universal phenomena in the out-of-equilibrium dynamics of quantum many-body systems 
has  { been intensified by efforts} at the interface of quantum information science and condensed matter physics to build quantum simulators \cite{altman2021quantum}, systems of hundreds of qubits which can realize  interesting many-body phases.
Analog quantum simulators, such as ultracold atom platforms \cite{lewenstein2007ultracold,schafer2020tools,gross2017quantum}, arrays of atoms with long-range interactions which are tuned by exciting into Rydberg states \cite{browaeys2020many}, and trapped atomic ions \cite{blatt2012quantum,lanyon2011universal} 
exert control over constituent degrees of freedom by tuning the Hamiltonian governing their interactions. {The possibility of highly coherent and controllable Hamiltonian dynamics}
has led to important questions about universality in the approach 
to thermal equilibrium \cite{gogolin2016equilibration, DAlessio-Rigol2016_review}, and situations in which quenched disorder can arrest thermalization entirely via the phenomenon of many-body localization (MBL)~\cite{nandkishore2015many,schreiber2015observation,abanin2019colloquium}. 
 
 Recently-developed \emph{digital quantum simulators} afford an even greater degree of control. The native mode of operation of these platforms, such as those formed of superconducting qubits \cite{wendin2017quantum,kjaergaard2020superconducting}, involves the discrete-time evolution of constituent qubits through the application of  \emph{unitary operations},  \emph{measurements}, and \emph{feedback}.  Harnessing these ingredients to control quantum many-body systems is a new goal of quantum condensed matter physics.  Successful implementation of these operations is also a stepping-stone towards the separate goal of eventually building a fault-tolerant quantum computer \cite{preskill2018quantum}.  The advent of digital quantum simulators has thus led to an important question for condensed matter physics: 
\begin{itemize}
    \item[] 
    What  collective quantum phenomena, or dynamical phases of matter,
   can emerge using operations -- such as unitary gates, measurements, and feedback -- which are native to quantum simulators?
\end{itemize}

This article is an introduction to simple discrete-time models for many-body dynamics that have allowed progress on some of these questions.
In these models, a lattice of spins (qubits) evolves through the application of local unitary gates and  measurements. 
This discrete-time structure  --- a so-called ``quantum circuit'' \cite{nielson2000quantum} --- is reminiscent of the ``Trotterization"  of a continuous-time Hamiltonian evolution, 
though the time-step here is not assumed to be infinitesimal, as each local operation will generally not be close to the identity (which also means that energy is not conserved).

A minimally-structured unitary quantum circuit which lacks any symmetries or other special properties will rapidly bring the system into a steady-state that is locally completely disordered, in the sense that local observables reproduce an infinite-temperature statistical ensemble.
The search for universal phenomena in this setting thus requires going beyond  traditional probes of quantum condensed matter {involving  correlations between specially-chosen local operators.}
Instead,  interesting universal features  of the evolving state can be usefully quantified with
 information-theoretic quantities  such as the entanglement entropy and the quantum mutual information,  measures of correlations that are  non-linear in the reduced density matrix for a subsystem. These observables are also natural in the minimal circuit setting since they are independent of the local choice of basis, while remaining sensitive to basic structural features of the dynamics such as locality and unitarity. 
What universal structures govern the behavior of such observables in a minimal quantum circuit dynamics?  What additional universal phenomena emerge in the presence of further structure and symmetries?

Focusing on these abstract measures of correlation is not an esoteric exercise; 
measures of entanglement  are of particular importance in light of recent developments in  condensed matter physics and quantum information science.
  First, closed quantum systems can reach a local, thermal equilibrium under purely unitary evolution, and understanding the locally irreversible nature of this process of \emph{thermalization} \cite{deutsch1991quantum,srednicki1994chaos,rigol2008thermalization} requires studying the production of quantum many-body entanglement.  
  Second, just as patterns of entanglement in equilibrium matter contain universal structures that are characteristic of phases and phase transitions \cite{wen2017colloquium,calabrese2009entanglement,eisert2010colloquium,swingle2010entanglement},  entanglement is also an organizing principle for out-of-equilibrium quantum matter.  It is interesting to compare universal patterns of entanglement that emerge in this setting with what we know to be possible in equilibrium. 
Finally, validating the performance of near-term quantum computers requires understanding and executing tasks which we know to be quantifiably hard to perform on a classical computer \cite{harrow2017quantum}.  
Quantum information-theoretic quantities
(e.g. measures of state and operator entanglement)
provide proxies for certain kinds of classical hardness, { so that --- in addition to shedding light on how to make classical algorithms more efficient --- they
can be used to pinpoint dynamical regimes where quantum simulation  has a genuine advantage.} 

Significant theoretical progress
 is possible by incorporating \emph{randomness} in the allowed local operations that form a quantum circuit.   Ensembles of random quantum circuits provide a theoretically tractable setting in which to understand universal out-of-equilibrium phenomena that also occur in more structured quantum many-body dynamics. 
 This is similar in spirit to the role of randomness in, for example, applications of random matrix theory to level statistics or mesoscopic transport \cite{brody1981random,beenakker1997random}. Loosely speaking, randomness in a quantum circuit allows for a classical description of the evolving entanglement structure  in a typical realization of the  quantum many-body dynamics \cite{nahum2017quantum,nahum2018operator,von2018operator,zhou2020entanglement}.
This result can be heuristically understood by noting that
basic observables (both simple correlation functions, and entanglement quantities that are non-linear in the density matrix)
are related to  probability amplitudes for evolving several copies of the original quantum many-body system into a particular final state. 
Randomness allows one to show that this evolution is quantum-mechanically incoherent.  The resulting ``classical'' 
statistical ensemble of Feynman trajectories 
of the multi-copy system exhibits universal properties which reflect
quantum correlations in the original system of interest. 
A key approach taken in this article will be to investigate the universal structures that emerge in these classical descriptions, and to use these as a foothold for understanding more structured quantum many-body evolution.  

{ 

We will discuss both unitary circuit dynamics,  and monitored dynamics in which the system's local degrees of freedom are  repeatedly measured by an external observer (a nonunitary operation).
One way to motivate unitary circuits is as models in which to get analytical control on questions about out-of-equilibrium dynamics and chaos that are also relevant to more conventional condensed matter models.
Monitored circuits, by contrast, show us that new dynamical universality classes are possible 
(even for dynamics with very little structure)
once we step outside the domain of unitary evolution. 
In particular we focus on a phase transition, from an entangled to a disentangled phase, induced by monitoring \cite{skinner2019measurement,li2018quantum}.
Unitary or weakly-monitored dynamics  generates complex, highly entangled wavefunctions, but sufficiently frequent measurement can trap the evolving wavefunction close to the space of product states. 
The phases and transitions can again be fruitfully mapped to an effective classical statistical mechanics model  \cite{jian2020measurement,
bao2020theory,li2021statistical,nahum2021measurement}. Monitored many-body systems have been the subject of much recent progress
\cite{li2019measurement,
chan2019unitary,
cao2018entanglement,
gullans2020dynamical,
gullans2020scalable,
jian2020measurement,
bao2020theory,
szyniszewski2019entanglement,
chen2020emergent,
zabalo2020critical,
zabalo2022operator,
turkeshi2020measurement,
li2021conformal,
lavasani2021measurement,
sang2021measurement,
lavasani2021topological,
vijay2020measurement,
alberton2021entanglement,
nahum2021measurement,
weinstein2022measurement,
buchhold2021effective,
jian2021measurement,
turkeshi2021measurement,
ippoliti2021entanglement,
lunt2020measurement,
lopez2020mean,
li2021entanglement,
roy2020measurement,
mcginley2021absolutely,
zhuang2021absolutely}.

We also direct the reader to a recent review \cite{potter2021entanglement} on entanglement dynamics in circuits, which gives a useful treatment of many of the same topics we discuss.}

This article is organized as follows. In Sec.~\ref{sec:models_motivation}, we review the fundamental building blocks of quantum circuits along with the two broad classes of quantum circuit dynamics that we study, and outline some basic physical properties of the circuits.
Sec.~\ref{sec:unitary_circuit} focuses on the first class, quantum circuits with local unitary gates.  We review a universal description of entanglement growth in this setting, and elucidate the structure of correlation functions.  We then explore circuits  with additional structure beyond locality and unitarity, such as charge-conserving circuits with a hydrodynamic mode, and Floquet circuits with discrete time-translation symmetry.
In Sec.~\ref{sec:monitored}, we explore a novel class of open quantum systems which are monitored by an observer who makes repeated measurements.  We describe ``hybrid" quantum circuits with both unitary gates and measurements, and a phase transition, driven by measurements, in  the entanglement structure of the quantum trajectories.  We explore properties of the phases and phase transitions in such circuits, and also discuss the role of additional structure such as discrete or continuous symmetries. Sec.~\ref{sec:Experiments}  discusses some experimental implications, and Sec.~\ref{sec:Outlook} is a brief outlook of open questions and future possible directions.

\section{Models and motivation}\label{sec:models_motivation}

We start this Section with
a pedagogical introduction to the essential building blocks of a quantum circuit at the level of one or two qubits,
to develop some intuition for the effect of these ingredients on correlations in a quantum many-body system. 
We  also introduce basis-independent measures of quantum correlations,
the entanglement entropy and the bipartite mutual information,
which we will use throughout this review. 
Then we define the 1+1D unitary (Sec.~\ref{sec:circuitdefns}) and monitored (Sec.~\ref{sec:measurementdefns}) circuits that we will consider. 
In Secs.~\ref{sec:useofmodels},~\ref{sec:qmclmappings} we briefly sketch what these circuits can be used to study and some of the structures allowing analytic calculations.

\subsection{Circuit building blocks and quantum entanglement}
\label{sec:buildingblocks}
 
Throughout this review, we will study $d$-dimensional quantum many-body systems composed of qubits (or more generally, $q$-level systems ``qudits")  which are arranged in a spatially local fashion.  The discrete time-evolution of such a quantum many-body system through the application of local operations defines a quantum circuit.   
We will restrict our attention to 
($i$) \emph{quantum gates} acting on a few nearby qubits and, in the second part of this review,  also
($ii$) local \emph{projective measurements}.   
Quantum gates are unitary transformations  acting on the qubits.
Projective measurements
 are observations which 
leave the measured degrees of freedom in a state with a definite eigenvalue of the measured operator.
Measurements are  inherently stochastic: identical measurements performed on multiple copies of the same wavefunction can yield different outcomes, which are distributed according to {Born's rule}.

We will make extensive use of the ``entanglement entropy''
(or rather entropies)
to quantify 
correlations in quantum states evolving under these ingredients. Consider a quantum many-body system composed of $N$ qubits and described by a wavefunction $\ket{\Psi}$, which is bipartitioned into a subset of spins $A$, with Hilbert space dimension $D_{A}$, and its complement  $\overline{A}$, with Hilbert space dimension $D_{\bar{A}}$. 
The entanglement entropy is a measure of the entanglement between $A$ and $\bar A$ and also a basis-independent measure of the correlations between these regions. It can be expressed in terms of the \emph{reduced density matrix} $\rho_A$, which encodes all expectation values of operators solely within region $A$.
This is given by tracing over the complementary subsystem $\overline{A}$:
$\rho_{A} \equiv \mathrm{Tr}_{\bar{A}}\ket{\Psi}\bra{\Psi}$. The eigenvalues  $\{\lambda_{i}\}$ of $\rho_A$ are  non-negative and sum to unity, $\sum_{i}\lambda_{i} = 1$, and as we will see below, the number of nonzero eigenvalues is also the number of terms required to write $\ket{\Psi}$ as a superposition of unentangled (product) states.

The {von Neumann entanglement entropy} is defined as 
\begin{align}
    S_{A} \equiv -\Tr\left(\rho_{A}\ln\rho_{A}\right).
\end{align}
We will often consider the $n^{\mathrm{th}}$ R\'{e}nyi entropy $S_{A}^{(n)} \equiv \ln\left(\mathrm{Tr}\rho_{A}^{n}\right)/(1-n)$ as  another measure of quantum entanglement, with the von Neumann entropy given by the limit $n\rightarrow 1$.

The entanglement entropy can also be used to define a basis-independent measure of correlations between subsystems $A$ and $B$ whose union is not necessarily the entire system. This ``mutual information'' is defined as 
\be
I_{AB}  = S_A + S_B - S_{A\cup B}.
\ee
The R\'{e}nyi and von Neumann entropies enjoy a number of important properties:   
\begin{enumerate}[leftmargin=1.28em,labelindent=16pt]
    \item {$S_{A}^{(n)}$ is insensitive to unitary transformations which act separately within $A$ or $\overline{A}$, such as local changes of basis in the quantum many-body system, since these leave the eigenvalues of the reduced density matrix unchanged.}
    \item $0 \le S_{A}^{(n)} \le \ln D$ where $D = \min(D_{A},D_{\bar{A}})$.  $S_{A}^{(n)}$ is zero if subsystem $A$ is \emph{pure} and may be described by a single wavefunction $\rho_{A} = \ket{\phi}\bra{\phi}$ and is maximized if the reduced density matrix for the smaller of the two subsystems is $I_{D}/D$ where $I_{D}$ is the $D\times D$ identity matrix (in this case we say that this subsystem is in a \emph{maximally-mixed state}).  In the former case, we will say that the two subsystems are \emph{disentangled}.
    \item Writing  $\rho_{A} = \sum_{i} p_{i}\ket{\phi_i}\bra{\phi_i}$ as a probabilistic mixture of orthonormal pure states (see below), with probabilities $p_i=\lambda_i$,
    the von Neumann entropy $S_{A} = -\sum_{i}p_{i}\ln p_{i}$ is the classical Shannon entropy of this distribution.
\end{enumerate}
The von Neumann entropy is often preferred to other R\'{e}nyi entropies as a measure of entanglement since it satisfies additional properties, such as sub-additivity,
strong sub-additivity, and concavity  which are natural if we wish to give  $S_A$ an information-theoretic interpretation \cite{nielson2000quantum}.\footnote{Sub-additivity means that if a region $A$ is decomposed into regions $A_1$ and $A_2$ then 
${S_{A}\leq S_{A_{1}}+S_{A_{2}}}$. { Heuristically, there is less ``uncertainty'' in $\rho_A$ than in the the combination of reduced states $\rho_{A_{1}}$ and $\rho_{A_{2}}$, since these are fully determined by $\rho_A$ \cite{ochs1975new}.}}

Heuristically, the entanglement entropy between complementary sets of spins $A$ and $\bar A$ quantifies (the logarithm of) the number of terms required to write the pure state $\ket{\psi}$ as a superposition of product states between $A$ and $\bar A$.  
Formally,  $\ket{\psi}$ may be written in the Schmidt form
 ${\ket{\psi} = \sum_{i} 
 \sqrt{\lambda_i}
 \ket{\phi_i}_{A} \ket{\chi_i}_{\bar A}}$, where $\ket{\phi_i}_{A}$ and $\ket{\chi_i}_{\bar A}$ are orthonormal sets of states in the two subsystems, and $i$ runs over at most ${\min(D_A, D_{\bar A})}$ values.
The distinct R\'enyi entropies correspond to distinct ``counts'' of the Schmidt values that discount smaller values of $\lambda_i$ to a greater or lesser extent.  

A key point is that low-entanglement states can be expressed (or approximated) by keeping a number of Schmidt states that is much smaller than $D$. A generalization of this idea to Matrix Product States (MPS) for 1D chains means that there is a direct relation between the entanglement of a 1D quantum state and the cost of storing it as an MPS \cite{schuch2008entropy}. Related ideas  apply also  to Matrix Product Operator representations of quantum operators.

\subsection{Quantum gates and measurements -- a two-qubit example}
\label{sec:12qubit}

To illustrate the effect of the basic building blocks of quantum circuits (quantum gates and measurements) on a quantum state and its entanglement, 
we consider 
examples involving only one or two qubits.

Starting with a single qubit,
recall that any Hermitian operator acting on the qubit may be written as a linear combination of the three  Pauli matrices and the identity operator,
\begin{align}
    X = \left(\begin{array}{cc} 0 & 1\\ 1 & 0\end{array}\right), \hspace{.2in} Y = \left(\begin{array}{cc} 0 & -i\\ i & 0\end{array}\right), \hspace{.2in} Z = \left(\begin{array}{cc} 1 & 0\\ 0 & -1\end{array}\right), 
    \hspace{.2in}
    I = \left(\begin{array}{cc} 1 & 0\\ 0 & 1\end{array}\right),
\end{align}
in the form ${\mathcal{O} = \sum_{\mathcal{S}\in \{I,X,Y,Z\}}c_{\mathcal{S}}\mathcal{S} = c_{I}I + c_{X}X + c_{Y}Y + c_{Z}Z}$, where the coefficients in this expansion are given by $c_{\mathcal{S}} = \frac{1}{2}\mathrm{Tr}(\mathcal{S}\mathcal{O})$, with the trace taken over the two-dimensional Hilbert space of the qubit.  
A unitary acting on the qubit can be written 
$u=\exp(- i  \sum_{\mathcal{S}\in \{I,X,Y,Z\}}h_{\mathcal{S}}\mathcal{S})$, where the coefficients $h_{\mathcal{S}}$ are real.
For a pair of qubits, any Hermitian operator $\mathcal{O}$ 
may be written as a sum of tensor products of these operators acting on the two qubits. 
(We will write $X_1=X\otimes I$, etc.)

Consider dynamics of 
a single spin that interleaves 
unitary gates and measurements.
If the spin is in a pure state,
this state is a point on the Bloch sphere, defined by the polarization vector ${(\<X\>,\<Y\>,\<Z\>)}$.
Unitary transformations rotate the state on the Bloch sphere.
A measurement, say of $Z$, causes a stochastic jump 
to  the North or the South pole, depending on the random measurement outcome 
(see below for an example in the two-qubit setting).
As a result of this randomness, an arbitrary sequence of unitaries and measurements gives a kind of random walk on the Bloch sphere.

For a single spin, all points on the Bloch sphere are equivalent (in the absence of a preferred local basis). 
But once we have more than one spin, we can distinguish wavefunctions according to their entanglement. 
As an illustration,  consider a simple process of generation and destruction of entanglement with two qubits. 
Begin with a state in which both spins are aligned in the Pauli $X$ basis $\ket{\psi} = \ket{\rightarrow\rightarrow}$, so that $X_{1}\ket{\psi} = \ket{\psi}$, $X_{2}\ket{\psi} = \ket{\psi}$, where $X_{i}$ denotes the Pauli operator $X$ acting on the $i$-th qubit.
 This state is \emph{disentangled} since it may be written as a tensor product of the wavefunction of each spin. We now apply the { ``controlled--Z''} gate acting on both qubits, CZ$_{12}\equiv \exp\left[i\displaystyle\frac{\pi}{4}(1 - Z_{1})(1-Z_{2})\right]$, so that the new wavefunction of the two-spin system is 
 \begin{align}\label{eq:wfn_gate_example}
     \ket{\phi} \equiv \mathrm{CZ}_{12}\ket{\psi} = \frac{1}{2}\Big[\ket{\rightarrow\rightarrow} + \ket{\rightarrow\leftarrow} + \ket{\leftarrow\rightarrow} - \ket{\leftarrow\leftarrow}\Big]
 \end{align}
 where $\ket{\leftarrow}$ is an eigenstate of the Pauli $X$ operator  with $X\ket{\leftarrow} = -\ket{\leftarrow}$. It is easily checked that the reduced density matrix for either spin is now maximally mixed, $\rho_{1,2} = I/2$, so that the entanglement entropy for each spin is $\ln 2$.  
 
 For most of this review, we will discuss generic local unitary gates. 
 The CZ unitary gate that we 
 applied above
  is in fact non-generic in two ways.  
  First, it is 
 diagonal in the Pauli $Z$ basis and therefore cannot generate quantum entanglement when acting on product states in this basis.  A generic two-qubit quantum gate acting on any product state of the qubits will tend to produce entanglement between them  (though not in general maximal entanglement). 
The second property is that $CZ$ is a ``Clifford'' gate, meaning that it has a simple action on Pauli matrices as we illustrate below. Although the generic gates that we will mostly discuss that do not have this special property, 
we describe it here briefly because circuits made of Clifford gates have the important property of being classically simulable (Sec.~\ref{sec:clifford}).
 
{The initial state $\ket{\psi}$ was ``stabilized by'' (invariant under)  the Pauli operators $X_1$ and $X_2$.}
An equivalent way to specify the evolved state $\ket{\phi}$
 is via the evolution of these
operators 
 under the action of the gate, which is given by $\mathrm{CZ}_{12}X_{1}\mathrm{CZ}_{12}^{\dagger} = X_{1}Z_{2}$,  $\mathrm{CZ}_{12}X_{2}\mathrm{CZ}_{12}^{\dagger} = Z_{1}X_{2}$. These evolved ``stabilizers'' uniquely specify the two-qubit wavefunction, since they satisfy  
 \begin{align}
     X_{1}Z_{2}\ket{\phi} = \ket{\phi} \hspace{.2in} Z_{1}X_{2}\ket{\phi} = \ket{\phi}
 \end{align}
Here, the Clifford property of CZ has ensured that the evolved stabilizers are simple products of Pauli operators. In contrast, a generic quantum gate will  evolve a Pauli operator into a sum of products of Pauli operators, as we will discuss in Sec.~\ref{sec:unitary:correlations}. 

Finally,  a \emph{projective measurement} of a spin will disentangle the measured spin.  Starting from the entangled state $\ket{\phi}$, a measurement of  $X_{1}$ will yield the outcomes $X_{1} = \pm 1$ with the respective probabilities
\begin{align}
    p_{\pm } &= \bra{\phi}
    \mathcal{P}_{\pm}
    \ket{\phi} = \frac{1}{2},
    &
    \mathcal{P}_{\pm} & \equiv\frac{1 \pm X_{1}}{2}, 
\end{align}
according to Born's rule.  After an observation of the outcome $X_{1} = +1$, the subsequent wavefunction of the two-qubit system will be given by
\begin{align}
\ket{\phi_{+}} \equiv \frac{\mathcal{P}_{+}\ket{\phi}}{\sqrt{p_{+}}} = \frac{1}{\sqrt{2}}\left[\ket{\rightarrow\rightarrow} + \ket{\rightarrow\leftarrow}\right] = \ket{\rightarrow\uparrow}
\end{align}
where 
$Z\ket{\uparrow} = \ket{\uparrow}$, so that the two qubits are now completely disentangled.  Similarly, an observation of the outcome $X_{1} = -1$ would have yielded the state $\ket{\phi_{-}} = \ket{\leftarrow\downarrow}$.  As a result, after the observation of any one outcome of the measurement, the two-qubit system remains in a pure state in which the constituent qubits are disentangled from each other. 
In contrast, the statistical mixture of pure states obtained by {averaging over} both measurement outcomes with their Born probabilities,
\begin{align}
\rho = \frac{1}{2}\ket{\phi_{+}}\bra{\phi_{+}} + \frac{1}{2}\ket{\phi_{-}}\bra{\phi_{-}}
\end{align}
--- which could describe the  density matrix of the  two-qubit system after an appropriate unitary interaction with an external bath --- is in a mixed state with entropy $-\mathrm{Tr}\rho\log\rho = \ln 2$, and the reduced state of each spin is maximally mixed due to the uncertainty in the measurement outcomes.

\begin{figure}[t]
\centering
\includegraphics[width=\columnwidth]{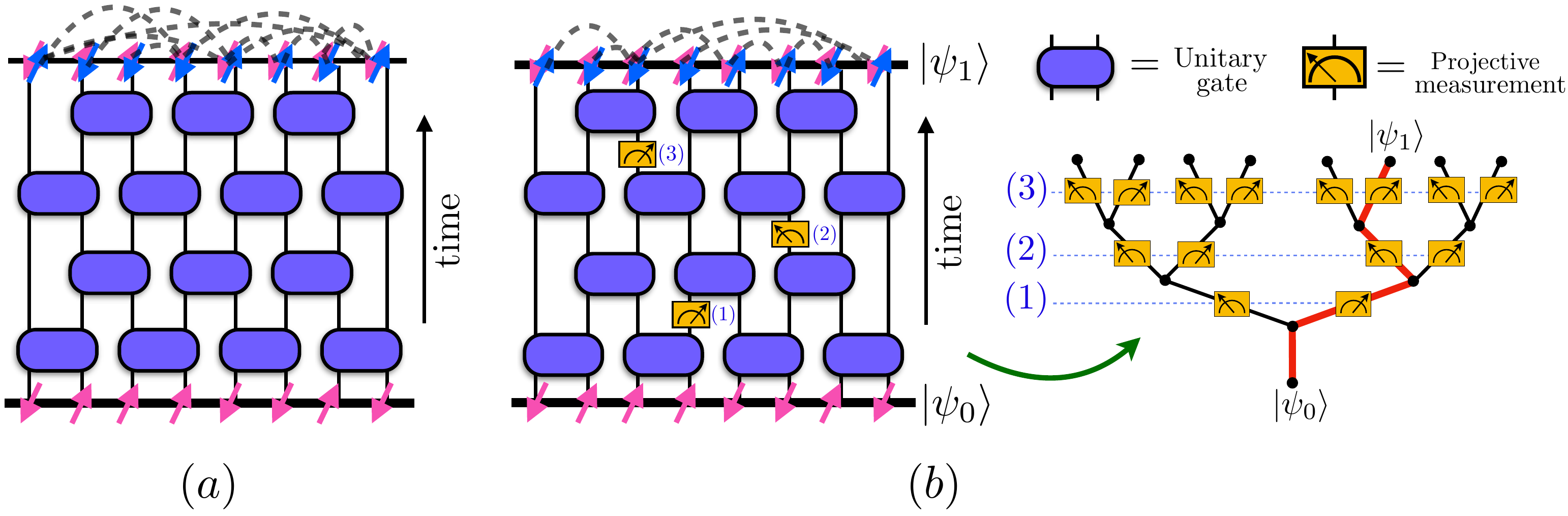}
\caption{A spacetime diagram of the two classes of quantum circuits considered in this review, with either $(a)$ a brickwork structure of two-site unitary gates or  $(b)$ unitary gates interspersed with local projective measurements.  A pure-state trajectory corresponding to a particular sequence of measurement outcomes is shown, though the inherently probabilistic nature of the measurement outcomes could yield other  trajectories, which are shown schematically in $(b)$. }
\label{fig:quantum_circuit}
\end{figure}

\subsection{Brickwork random unitary circuit}
\label{sec:circuitdefns}
  
We now turn to unitary  dynamics for a chain of $L$ qubits.
In the traditional many-body setting, the starting point would be a Hamiltonian specifying evolution in continuous time. 
Quantum circuits (used extensively in quantum information \cite{nielson2000quantum}) abstract away from this and instead specify evolution in discrete time ${t\in \mathbb{Z}}$. For each timestep there is a unitary evolution operator ${U_t = U(t;t-1)}$,
under which a pure state (in the $2^L$-dimensional many-body Hilbert space) evolves as ${\ket{\psi_t} = U_t \ket{\psi_{t-1}}}$.
 $U_t$ is taken to be a tensor product of local unitary gates that act on pairs of spins, with the alternating structure shown in  Fig.~\ref{fig:quantum_circuit}:
\ba\label{eq:circuitdefn}
U(t;0)& =U_t\ldots U_2 U_1, \\ 
U_\tau & =
\left\{
\begin{array}{ll}
 {\otimes}_{x \in \text{odd bonds}}    \,\,\,  u_{\tau,x} & \text{if $\tau$ is odd},   \\
 {\otimes}_{x \in \text{even bonds}}  \,\,\,  u_{\tau,x} & \text{if $\tau$ is even}.
\end{array}
\right.
\end{align}
We have labelled bonds of the lattice by integers $x$. We leave the boundary conditions unspecified for now.
Here $u_{\tau,x}$ is a local gate, acting on the $4\times 4$ Hilbert space for a pair of spins, which is applied (in timestep $\tau$) to the two spins connected by bond $x$.

Fig.~\ref{fig:quantum_circuit}a can be thought of in several ways. First, it is a space-time diagram specifying which local interactions are ``switched on'' during which timesteps. 
Second, it is also a tensor network, specifying how to build the full many-body unitary $U(t;0)$ from four-legged tensors $u_{\tau,x}$ by contracting spin indices on all of the bonds. This contraction means  summing over $\uparrow$ and $\downarrow$ on each bond (if we use, say, the  $Z$ basis).
Performing this 
contraction, with the spin 
indices at the top and bottom of the 
circuit fixed to ${\{a_1\ldots a_L\}}$ 
and  ${\{b_1\ldots b_L\}}$ respectively, 
yields the
amplitude 
${\langle a_1\ldots a_L |U(t;0)| b_1\ldots b_L \rangle}$
to propagate between given initial and final states.
Physically, the index sums represent the sum over Feynman histories of the spin chain, so, with appropriate boundary conditions at the initial and final time, a final way of viewing the circuit is as a discrete real-time path integral. 

The model generalizes immediately to the case where the local degrees of freedom are ``quqits" with local Hilbert space dimension $q\geq 2$, and the local gates lie in the unitary group ${\mathrm{U}(q^2)}$.

We start with minimally structured circuits, in which  every local gate $u_{x,\tau}$ is drawn randomly \textit{and independently of all the others} from the uniform distribution on the unitary group  ${\mathrm{U}(q^2)}$. This uniform (Haar) distribution is defined by the invariance of all averages involving the random unitary $u_{x,\tau}$ under both left and right unitary rotations, ${u_{x,\tau}\rightarrow v u_{x,\tau} w}$, for any choices of ${\mathrm{U}(q^2)}$ matrices $v$ and $w$.
We will discuss these averages  \cite{weingarten1978asymptotic,collins_integration_2006} in Sec.~\ref{sec:unitary_circuit}.  

Random circuits can be used to model many kinds of dynamical process, and in many situations we will wish to impose more structure or symmetry than in the model above. 
(We note that even the ``minimal'' brickwork circuit above posseses one basic structure, which is spatial locality of the interactions --- all-to-all-coupled circuits, in which spatial locality is relaxed, are natural in quantum information and as toy models for black holes \cite{hayden2007black, sekino2008fast, lashkari_towards_2013, shenker2015stringy}, and we will also touch on them later in the monitored setting.)
In Secs.~\ref{sec:u1},~\ref{sec:floquet} we will review the incorporation of  global symmetries  \cite{khemani2018operator,rakovszky2018diffusive} or time translation symmetry \cite{chan2018spectral,chan2018solution} into the random circuit. Alternative choices are also possible for the distribution of unitaries or for the  geometry of the circuit
(in Sec.~\ref{sec:mincut} we will   consider the case where the regular brickwork of Fig.~\ref{fig:quantum_circuit} is replaced with a random spacetime structure). For the moment, however, we view the dynamics in a given realization of the circuit --- i.e. with particular choices of the random gates $u_{\tau,x}$ --- as an example of a ``chaotic'' quantum evolution, and compare it with more conventional nonintegrable many-body systems.

The circuit  has no conservation laws (not even energy conservation).   But the usual lore of local equilibration still applies.  Under the evolution, an arbitrary initial state (pure or mixed) eventually equilibrates locally 
to the Gibbs state \cite{nandkishore2015many}. As there are no conserved quantities here, this is simply the featureless infinite temperature state 
\cite{lazarides2014equilibrium,d2014long,ponte2015periodically}, whose entropy density ${\seq=\ln q}$ is set by the local Hilbert space dimension. The reduced density matrix of a subregion of a pure state obtains this entropy in the form of entanglement with the rest of the system, as we review in Sec.~\ref{sec:unitary:entanglement}. 

There is no notion of a ground state or of elementary excitations above a ground state.
Indeed, since the circuit does not even have  a discrete time-translation symmetry (except ``on average''), the eigenstates of  $U(t;0)$ are time-dependent, and unlikely to be a useful starting point for computing observables. This is quite different from standard many-body systems (e.g. Fermi liquids) at asymptotically \textit{low} temperature, when elementary excitations become long-lived and provide a useful description of the dynamics. But the random circuits in Eq.~\ref{eq:circuitdefn} (and more structured extensions of them) have proven to be useful  models for various phenomena in nonintegrable dynamics at higher temperature, when the relevant timescales are much longer than the timescale for nonintegrable scattering of quasiparticles (so that quasiparticles stop being a useful language).
 We discuss what we can hope to study using these models in Sec.~\ref{sec:useofmodels} and Sec.~\ref{sec:unitary_circuit}.

\subsection{Measurements and trajectories}
\label{sec:measurementdefns}

We now review some points about measurements that will reappear when we discuss monitored circuits in Sec.~\ref{sec:monitored}.

Measurements affect the state, and if the number of local measurements is extensive in the spacetime volume, they fundamentally alter the dynamics (Sec.~\ref{sec:monitored}).
The circuit is a  simple setting for studying how this happens, and for exploring how we should define dynamical phases and dynamical universality classes in ``monitored'' many-body systems. A basic concept is the quantum trajectory. 

Imagine that an experimentalist makes a sequence of $M\geq 1$ local measurements, at various locations and times, during the circuit evolution of a pure state $\ket{\psi}$. For concreteness let these be projective measurements of individual spins in the $\sigma_z$ basis. 
If spin $i$ is measured, the state undergoes the stochastic evolution
\be\notag
\ket{\psi} \rightarrow
\left\{
\begin{array}{ccc}
\mathcal{P}_{i\uparrow} \ket{\psi}/ \sqrt{p_{i\uparrow}}   &  \text{with  probability $p_{i\uparrow}=\bra{\psi}\hspace{-0.5mm} \mathcal{P}_{i\uparrow}\hspace{-0.5mm}  \ket{\psi}$}  \\
\mathcal{P}_{i\downarrow} \ket{\psi} / \sqrt{p_{i\downarrow}}   & \text{with  probability $p_{i\downarrow}=\bra{\psi}\hspace{-0.5mm} \mathcal{P}_{i\downarrow}\hspace{-0.5mm}  \ket{\psi}$} 
\end{array}
\right.
\ee
(Born's rule), where $\mathcal{P}_{im}$ projects spin $i$ onto $Z_i = m$. 

In a given ``run'' of this experiment, the experimentalist will obtain a random sequence ${{\bf m} = (m_1, \ldots, m_M)}$ of measurement outcomes, with ${m_\alpha = \uparrow, \downarrow}$. Note that this stochasticity should not be confused with the randomness in the unitaries.
For example, we can imagine that the sequence of unitaries $\{ u_{x,t}\}$ has been fixed in advance, and the experiment is repeated several times using the same unitaries. Distinct runs will  still yield distinct ${\bf m}$ in general.

A given measurement record ${\bf m}$, and the associated evolving state $\ket{\psi_{\bf m}(t)}$, defines a trajectory. Note that, so long as our imagined experimentalist keeps a record of the measurement outcomes ${\bf m}$, the measurement events do not introduce any classical uncertainty about the state --- it remains pure. This is \textit{unlike} the interaction of an open system with a bath, where information loss to the bath forces us to work with a mixed state. (But there are formal connections between the two settings \cite{plenio1998quantum,gardiner2004quantum,
daley2014quantum} that we discuss in Sec.~\ref{sec:trajectoryvsmixed}.)

Another basic point is that the repeated measurements do not simply read out a preexisting unitary dynamics: they yield a new dynamics, which is a kind of random walk through Hilbert space.
If our system was only a single spin, this would be a trajectory on the Bloch sphere, with measurements causing ``quantum jumps'' 
\cite{plenio1998quantum,gardiner2004quantum,
daley2014quantum,nagourney1986shelved,sauter1986observation,
 bergquist1986observation} 
 to the North or South pole (Sec.~\ref{sec:12qubit}).
In the many-body case the problem is richer, because having \textit{local} degrees of freedom means that not all points in Hilbert space are equivalent: states can be distinguished by their entanglement structure and their complexity.
A local measurement disentangles spin $i$ from all of the others. This effect competes with the spreading of correlations by the unitaries, leading to a phase transition that we review in Sec.~\ref{sec:monitored} \cite{skinner2019measurement,li2018quantum,li2019measurement}.

We will focus on the case where the measurements occur at a finite rate per degree of freedom. 
A simple choice is to let each spin be measured with probability $p$ in a given timestep, i.e. to scatter measurement events through spacetime with probability $p$. This gives us a tuning parameter $p$ for the strength of monitoring.

Let us formalize the monitored evolution over a time interval $[0,t]$ as a circuit. Without measurements, we would have 
a unitary circuit $U=U(t;0)$.  With measurements, we can define a nonunitary circuit, $K_{\bf m}$, for any given sequence of outcomes ${\bf m}$. This circuit is obtained from $U$ by introducing projection operators, on bonds of the tensor network, at the  spacetime locations of the measurements. Repeatedly applying Born's rule shows that the probability of a trajectory ${\bf m}$ is ${p_{\bf m} = \bra{\psi(0)}K_{\bf m}^\dag K_{\bf m} \ket{\psi(0)}}$, and the final state is
${\ket{\psi_{\bf m}(t)}=K_{\bf m} \ket{\psi(0)} /\sqrt{p_{\bf m}}}$. 
It is also enlightening to study the evolution of mixed states \cite{gullans2020dynamical}: the outcome sequence ${\bf m}$ has probability ${p_{\bf m} =\Tr K_{\bf m} \rho(0)K_{\bf m}^\dag}$, and gives the final state ${\rho_{\bf m}(t) = K_{\bf m} \rho(0) K_{\bf m}^\dag/p_{\bf m}}$.

This measurement process may appear abstract. But it is a step into a rich landscape of ``monitored'' many-body quantum systems, with  a wide range of different phases and phase transitions that are beginning to be explored \cite{skinner2019measurement,li2017measuring,chan2019unitary,cao2018entanglement,lavasani2021measurement,sang2021measurement,lavasani2021topological,vijay2020measurement,alberton2021entanglement,nahum2021measurement,weinstein2022measurement,buchhold2021effective,jian2021measurement,turkeshi2021measurement,ippoliti2021entanglement,lunt2020measurement,lopez2020mean,li2021entanglement}. 
The model also reveals a phase transition in the computational complexity, for a classical computer, of simulating various kinds of open or monitored quantum dynamics. Finally it may be a toy model for certain process in quantum information processing or encoding \cite{yoshida2021decoding,li2021statistical,fan2020selforganized,choi2020quantum,gullans2020dynamical,gullans2021quantum}. We discuss these topics further in Sec.~\ref{sec:monitored}.

  \subsection{What can we hope to do with these models?}
  \label{sec:useofmodels}

One use of the  1+1D  circuits without measurements is as tractable models for nonequilibrium dynamics, thermalization, and entanglement generation in local many-body systems \cite{oliveira_generic_2007,
znidaric_exact_2008,
hamma2012quantum,  
nahum2017quantum, nahum2018operator,von2018operator,
 zhou2019emergent}.
All-to-all random circuits (in which there is no notion of distance and any qubit can interact with any other) have also been used as models for scrambling in black holes \cite{hayden2007black, sekino2008fast, lashkari_towards_2013, shenker2015stringy}.
In these contexts, randomness is a tool to promote solvability.  Studying a specific nonintegrable model is hard. But in the random circuit it is often possible to obtain exact results for averages over the ensemble (or for typical instances). This is in the general spirit of other uses of randomness such as random matrix theory \cite{haake1991quantum} (or even coding theory \cite{shannon1948mathematical}, or more recent models like the Sachdev-Ye-Kitaev model \cite{sachdev1993gapless,kitaevtalk}).

Even in the absence of conservation laws (so that there are no hydrodynamic ``slow modes'') there is long-timescale dynamics associated with the spreading of quantum information that can be studied using the minimal random circuits above. 
Having a solvable model may allow us to identify coarse-grained structures that govern a broader universality class of systems. We will describe examples of emergent structures in the following sections. The solvable setting also gives hints as to how to do quantitative computations in more realistic models. 

If the additional symmetry of time-translation invariance is added to the circuits \cite{chan2018spectral,kos2018many}, then   traditional diagnostics of quantum chaos, using the eigenvalue spectrum of the time-evolution operator,  may be studied (Sec.~\ref{sec:floquet}).
Conservation laws can be added, in order to explore the emergence of hydrodynamics \cite{khemani2018operator, rakovszky2018diffusive}.
The circuit architecture also allows  structures that are not available with a fixed Hamiltonian:
one can for example impose a duality between the space and time directions \cite{bertini2019exact} (Sec.~\ref{sec:dualunitary}),
or restrict the set of allowed unitary gates so that the dynamics is classically simulable \cite{Gottesman_Knill} (Sec.~\ref{sec:clifford}).

The circuits are a natural setting for adding further ingredients to the dynamics, potentially influenced by ideas from quantum computing or NISQ devices. There is a lot of space to explore between traditional models of many-body systems (which are left to evolve with a fixed Hamiltonian) and the highly structured evolutions, with unitaries and measurements, relevant to quantum information.
Random circuits are also important in quantum information  as ingredients of proofs or algorithms \cite{brandao2016local},
and for benchmarking 
(non-random) quantum circuits \cite{liu2021benchmarking,cross2019validating}. Since sampling the output of a sufficiently deep, random quantum circuit is believed to be prohibitively difficult for a classical computer \cite{bouland2019complexity}, an ensemble of random circuits has recently been implemented experimentally in an attempt to demonstrate ``quantum supremacy" \cite{arute2019quantum}, leading to further interest in the complexity-theoretic aspects of random circuit dynamics \cite{bouland2019complexity,liu2021benchmarking,aaronson2019classical}.

\subsection{Quantum-classical mappings}
\label{sec:qmclmappings}

A  theme of the following will be  mappings between real-time quantum dynamics and effective classical statistical mechanics models.  In simple limits (e.g. large local Hilbert space dimension $q$) 
some observables  reduce to ``classical''  geometrical properties of the circuit
(related for example to its ``min-cut'' structure or its lightcone structure, as we discuss in later sections)
and these limits already give some insight. But it is possible to relate  the discrete path integral defined by the circuit ${U=U(t;0)}$ to effective ``classical'' ensembles much more generally.

More precisely, most physical observables are expressed as ``multi-sheet'' path integrals \cite{calabrese2009entanglement,kamenev2011field,aleiner2016microscopic}, or rather multi-layer circuits (an example is shown in Fig.~\ref{fig:haar_avg}).
Even  simple expectation values such as ${\langle\mathcal{O}(t)\rangle=\bra{\psi(0)}U(t,0)^\dag  \mathcal{O} U(t,0)\ket{\psi(0)}}$ involve the circuit and its complex conjugate: these can be represented with a doubled  circuit in which $U$ and $U^*$ form two stacked  layers.
Entropies are nonlinear in the density matrix, and require larger numbers of  layers.
A key feature of the random circuits is that these multi-layer circuits simplify --- only a drastically reduced set of spacetime configurations survive averaging, and in simple cases the resulting  configuration sum can be mapped to an effective classical partition function.

We will touch on two structures. First, for small numbers of layers circuit averages can be  mapped to classical Markov processes \cite{oliveira_generic_2007,dahlsten_emergence_2007,znidaric_exact_2008,harrow_random_2009,hamma2012quantum,zanardi2014local,nahum2018operator,von2018operator,khemani2018operator},  describing the incoherent (dephased)  evolution of quantum operators (Sec.~\ref{sec:unitary:correlations}, Sec.~\ref{sec:u1}). 
Second, for a general number of layers, it is possible to make a mapping to an effective ``classical'' lattice magnet, for degrees of freedom labelling pairings between layers of the circuit (Secs.~\ref{sec:entanglement:pairings},~\ref{sec:measurementdwsec}). A key role is played by domain walls in these effective magnets \cite{nahum2018operator,hayden2016holographic,zhou2019emergent, zhou_operator_2020,vasseur2019entanglement,jian2020measurement,bao2020theory,chan2018solution,chan2018spectral,hunter2019unitary,liu2021entanglement}. In 1+1D these domain walls are paths, and many basic calculations of entropies and correlators reduce to simple   random walk problems.  We will also touch on how to generalize these ideas beyond random circuits.

\section{Unitary circuit dynamics}\label{sec:unitary_circuit}

In this Section we focus on local unitary dynamics, starting with the minimally structured case before moving on to circuits with additional symmetries or invariance properties:
 hydrodynamic modes (Sec.~\ref{sec:u1}), time-translation symmetry (Sec.~\ref{sec:floquet}), space-time rotation symmetry (Sec.~\ref{sec:dualunitary}), or classical simulability (Sec.~\ref{sec:dualunitary}).

A useful way to quantify correlations in the minimally-structured models of Sec.~\ref{sec:circuitdefns} is with entanglement entropies of spatial regions  (and mutual informations between them),
because these information-theoretic quantities can be formulated without reference to any structure except locality.
We start with the entanglement, which is useful to understand the spreading of quantum information on large lengthscales,  before discussing more conventional local correlation functions (Sec.~\ref{sec:unitary:correlations}).
At the same time, we introduce some of the key structures underlying the analytical  tractability of the circuits.

A basic point is that locality of the gates imposes a bound on the speed at which any information can spread \cite{lieb1972finite}. 
The geometry of the circuit immediately implies an upper bound of unity for this speed,  but the characteristic ``butterfly velocity'' $v_B$ 
\cite{roberts_localized_2015_1,roberts_liebrobinson_2016}
for spreading of operators is typically below this geometric bound (Sec.~\ref{sec:unitary:correlations}). Similarly the structure of the circuit imposes a bound on entanglement growth that is discussed in Sec.~\ref{sec:mincut}.

\subsection{The entanglement membrane and the pairing order parameter}
\label{sec:unitary:entanglement}

Unitary evolution generically transforms weakly entangled pure states, which are atypical in Hilbert space \cite{lubkin1978entropy,page1993information,nadal2011statistical}, into volume-law states 
\cite{calabrese2005evolution,kim2013ballistic,liu2014entanglement,nahum2017quantum,mezei2017entanglement,
jonay2018coarse}. This entangling process underlies pure state thermalization, and sets  limits on the power of MPS and MPO simulation algorithms \cite{schuch2008entropy}. 

The entangling process is also a simple setting for introducing
a basic structure underlying various calculations in random circuits, random Floquet circuits, and the monitored circuits discussed later in this paper, which involves an order parameter for ``pairing'' between Feynman trajectories in a path integral for multiple forward and backward paths \cite{nahum2018operator,zhou2020entanglement,chan2018solution,jian2020measurement,bao2020theory}. 
We discuss this in  Sec.~\ref{sec:entanglement:pairings}.  

We start by sketching the continuum picture for entanglement production  that arises from the lattice calculations (which we discuss subsequently).
Random circuits led to a coarse-grained description in which a cost is assigned to surfaces in spacetime \cite{jonay2018coarse, nahum2017quantum}. This theory has a broader application to chaotic systems \cite{mezei2018membrane,chan2018solution, zhou2020entanglement}, including holographic field theories where remarkably explicit computations are possible~\cite{mezei2018membrane, mezei2020exploring}. 
It is also possible to discuss the entanglement of operators, as opposed to states, but we stick here to the example of entanglement generation in a state.

\subsubsection{Membrane picture}\label{sec:membrane}

Consider the growth of von Neumann entanglement entropy $S_A(t)$ for a spatial region $A$ in an infinite $d$-dimensional system, after quenching from a product state. 
Let us take $A$ to have a finite boundary area  (however in 1D $A$ could be semi-infinite).
For generic local dynamics, entanglement is initially (i.e. on timescales 
small compared to the size of the region, though
large compared to microscopic timescales) generated at a rate proportional to the area of the boundary of $A$ \cite{calabrese2009entanglement, kim2013ballistic,liu2014entanglement,dahlsten_emergence_2007}. If $A$ is finite, then at asymptotically late times thermalization of the reduced density matrix $\rho_A$ requires $S_A(t)$ to saturate to the thermal entropy  ${S_A(\infty)=\seq  V_A}$ (where $V_A$ is the volume of $A$).

In the membrane picture (and at leading order when length and timescales are large) $S_A$ is given  by minimizing an effective ``free energy'', or entanglement cost, for a membrane. 
This membrane is a $d$-dimensional surface within the ${d+1}$-dimensional spacetime slab ${t'\in [0,t]}$. 
{ At the final time (``top'') surface ${t'=t}$, this membrane is anchored to the boundary $\partial A$ between region $A$ and its complement $\bar A$, and the membrane separates  these two regions of the top surface, in the sense  that any path from one to the other that passes through the bulk of the spacetime, must intersect  the membrane.}
At least for small enough $t$, the membrane spans the full time interval $[0,t]$,  so has an   effective free energy that grows with $t$.  
As a result, 
 the entanglement
$S_A$ grows linearly with $t$
 and at a rate
proportional to $|\partial A|$, reflecting
the generation of correlations  over
a ballistically growing
 lengthscale  \cite{calabrese2005evolution,kim2013ballistic,liu2014entanglement}.
If $A$ is finite, then at some time, proportional to  the linear size of $A$, there is a discontinuous   transition to an optimal membrane configuration that closes off in the bulk,  whose cost equals the late-time entropy~${S_A(\infty)=\seq  V_A}$.

The key quantity that we need to know to perform this minimization is a model-dependent ``membrane tension'' $\mathcal{E}(v)$. This tension depends on the  local slope of the membrane with respect to the time axis, which can be parameterized with a velocity $v$. 
 $\mathcal{E}(v)$ is highly constrained by causality and unitarity (note for example that we must reproduce the correct late-time entropy.)

The simplest case is the entanglement $S_y(t)$ of a semi-infinite region $(-\infty, y]$ in an infinite 1D chain. The membrane is then a trajectory $x_{t'}$ with $x_t=y$ and $x_0$ free. For a class of initial states,
\be
S_{y}(t) = \seq \, \min \, \lf \int_0^t \dd t' \mathcal{E}(\dot x) 
+ S_{x_0}\hspace{-0.4mm}(0)
\ri,
\ee
where the minimization is over trajectories,
{ and $\seq$ is the equilibrium entropy density ($\ln q$ in the random circuit).} This may also be written
\be\label{eq:entanglementrateeq}
\partial_t S_y = \seq \, \Gamma\lf \partial_y S_y \ri,
\ee
where the entanglement production rate $\Gamma$ is a Legendre transform of $\mathcal{E}$.\footnote{We have assumed the dynamics to be homogeneous after coarse-graining
as in the circuits discussed in Sec.~\ref{sec:circuitdefns} and below. 
Otherwise, $\mathcal{E}$ and $\Gamma$ may have explicit dependence on space or time: an example is a system with spatial  ``weak links'' \cite{nahum2018dynamics}.
In general $\mathcal{E}$ will also depend on any hydrodynamic densities that are present and on a thermodynamic entropy current if conservation laws allow one \cite{mezei2020exploring,rakovszky2019entanglement,gong2022coarse}. }

For a parity-symmetric system $\mathcal{E}(v)$ is a convex function with its minimum at $v=0$, so for a product state quench the optimal path is straight (${\dot x = 0}$).
By definition $\mathcal{E}(0)=v_E$ gives the entanglement growth rate in this setting: { 
$S_y(t) = \seq v_E t$.}
The line tension $\mathcal{E}(v)$ also contains information about the operator spreading velocity $v_B$ discussed in Sec.~\ref{sec:unitary:correlations}, though this is in general distinct from $v_E$: we may argue via causality that $\mathcal{E}(v_B)=v_B$ and $\mathcal{E}'(v_B) =1$ \cite{jonay2018coarse}.

\subsubsection{Minimal cut}\label{sec:mincut}

\begin{figure}[t]
\includegraphics[width=\columnwidth]{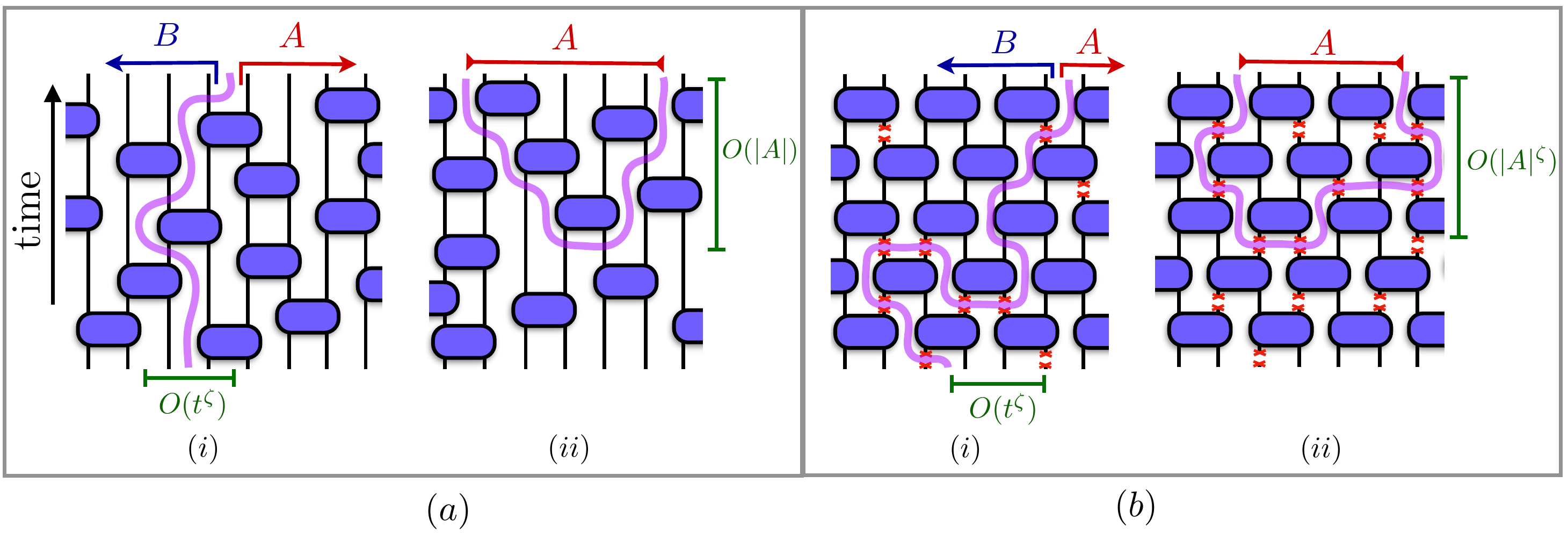}
\caption{Minimal cuts for (a) unitary dynamics with  geometric randomness in the applied quantum gates and  (b) monitored dynamics, with randomly-applied single-site projective measurements, which are discussed in Sec.~\ref{sec:monitored}.
In (b), the locations of the projective measurements are indicated by bonds with an ``x", which do not contribute to the cost of a minimal cut. In both kinds of dynamics, ($i$) shows the minimal cut corresponding to the growing entanglement of a subsystem, while ($ii$) shows the cut corresponding to the saturated entanglement of a finite sub-region. In the unitary case, the minimal cut may be taken to be strictly ``causal'' in the sense defined in Sec. \ref{sec:mincut}. The transverse wandering of the minimal cut relative to its fixed endpoint(s) is governed by the  \emph{wandering exponent} $\zeta = 2/3$ of the directed polymer in a random environment (DPRE), as indicated in the relevant panels, which is connected with universal fluctuations of the entanglement growth (see Sec. \ref{sec:mincut}) as well as universal sub-leading corrections to the entanglement entropy in the volume-law phase of the monitored dynamics (see Sec. \ref{sec:measurementdwsec}).
}
\label{fig:min_cut}
\end{figure}

The simplest version of the membrane appears in the limit 
${q\rightarrow \infty}$, where it can be related to the idea of a ``minimal cut''.
The minimal cut is a way of bounding entanglement in any tensor network 
\cite{swingle2012entanglement,chamon2012virtual,perez2007matrix,verstraete2008matrix,casini2016spread}. 
We temporarily exchange the brickwork circuit of Sec.~\ref{sec:circuitdefns} for one with a random spacetime structure: gates are applied to bonds at random times, in a Poisson fashion, at rate 1 per bond: the reason for this is mentioned below.
 A cartoon of the resulting circuit is shown in Fig. \ref{fig:min_cut}a.

When $q\rightarrow\infty$, the entanglement $S_y/\seq$ is given exactly by the cost of a minimal directed cut through this random network \cite{nahum2017quantum} --- see Fig. \ref{fig:min_cut}a.  
This cut is the solution to a nontrivial random optimization problem, so its exact cost depends on the circuit realization. 
Nevertheless it has a well-defined \textit{deterministic} line tension at large scales, telling us how many bonds we need to cut, per unit time, in order to bisect the circuit at a given angle. This is ${\mathcal{E}(v) = \f{1}{2} (1+v^2)}$ for ${|v|\leq 1}$, and ${\mathcal{E}(v) =1}$ for ${|v|\geq1}$.  The operator spreading speed may be argued to be $v_B=1$ at large $q$ 
[this is related to the average growth rate of the ``lightcone'' that is appears in Fig.~\ref{fig:min_cut}a(ii)]
so the constraints above are satisfied.
The entanglement production rate  $\Gamma$ in Eq.~\ref{eq:entanglementrateeq} is  ${\Gamma(s)=\f{1}{2}(1-(s/\seq)^2)}$, so that $\partial_t S$ is largest if $\partial_x S$ is zero.

$\mathcal{E}(v)$ sets the leading, deterministic growth of $S_y(t)$ at large $t$. 
This picture also reveals subleading structure arising from \textit{randomness} in the circuit. 
At the moment this randomness is the geometrical randomness in the circuit, but we will see below that randomness in the unitaries has a similar effect, even in the geometrically regular brickwork circuit, when $q$ is finite.

Finding the optimal cut is a version of the ``directed polymer in a random medium'' problem,   with  nontrivial critical exponents \cite{huse1985huse,
kardar1986dynamic}. 
In the quench from the product state, ${S_y(t)}$ has the Kardar-Parisi-Zhang (KPZ) scaling ${S_y(t) = \seq \lf v_E t + b t^{1/3} \chi_{y,t} \ri}$. The exponent  $1/3$ is of the subleading term is universal.  $\chi_{y,t}$ is a random variable, of order 1 size,  that depends on the realization  \cite{halpin2015kpz,corwin2012kardar,kriecherbauer2010pedestrian}.  These fluctuations may also be understood in the language of Eq.~\ref{eq:entanglementrateeq}, which, after including subleading terms, including noise, and expanding $\Gamma$ to quadratic order,  becomes the KPZ stochastic equation for the entropy profile~$S_y(t)$. 

 The minimal cut also describes the saturation of the entanglement entropy of a finite region at late times.  In (1+1)-dimensions, the entanglement of a region of finite width $\ell$ (in an infinite system) will grow until a time ${t_{\star} = \ell/2v_{E}}$, at which point the entanglement of the region will have reached its maximum value ${S_{\ell}(t>t_{\star}) = \seq \ell}$.  At times $t<t_{\star}$, the minimal cut for $S_{\ell}(t)$ will consist of two curves, one for each endpoint of the region, which are ``vertical'' (oriented in the time direction).  At times ${t> t_{\star}}$, the minimal cut consists of a single curve connecting the two endpoints of the region which intersects $\ell$ bonds of the circuit.  This curve is not unique, but  it  can be chosen to be oriented along the  lightcone of the quantum circuit as in Fig.~\ref{fig:min_cut}a. 
 In the formalism of Sec.~\ref{sec:entanglement:pairings}, the analogous membrane has a definite coarse-grained $V$ shape, made up of two segments that travel at speed $\pm v_B$.

The minimal cut is an exact description of the entanglement entropy only when ${q=\infty}$. However, many aspects of the phenomenology survive at finite $q$. Next we discuss a more generic approach to deriving the membrane.

\subsubsection{The pairing order parameter}
\label{sec:entanglement:pairings}

The R\'enyi entropies are examples of observables that can be expressed in terms of ``multi-layer'' circuits, built by stacking copies of $U(t)$ and $U(t)^*$ 
and attaching appropriate boundary conditions to the initial and final-time bonds: the $n^{\mathrm{th}}$ R\'{e}nyi entropy  involves $n$ copies each  of $U(t)$ and $U(t)^*$, as discussed below for the example of $S_2$. 
Formally, such a stack describes unitary evolution going on in parallel in several copies of the spin chain.
This evolution may be formulated in terms of Feynman trajectories for each copy (the sum over trajectories is equivalent to the contraction of the tensor network, see Sec.~\ref{sec:circuitdefns}).
A key quantity in this setting is an ``order parameter'' labelling pairing between Feynman trajectories \cite{nahum2018operator, von2018operator, zhou2019emergent, chan2018solution, zhou2020entanglement, hunter2019unitary}. 
Heuristically, phase cancelation is avoided if the trajectory for each ``forward'' [i.e. $U(t)$] layer is locally similar to that for a paired ``backward'' [i.e. $U(t)^*$] layer, and there are several ways to choose the pattern of pairing.
This intuition can be made precise in the brickwork circuit of Sec.~\ref{sec:circuitdefns}.

\begin{figure}[t]
\includegraphics[width=\columnwidth]{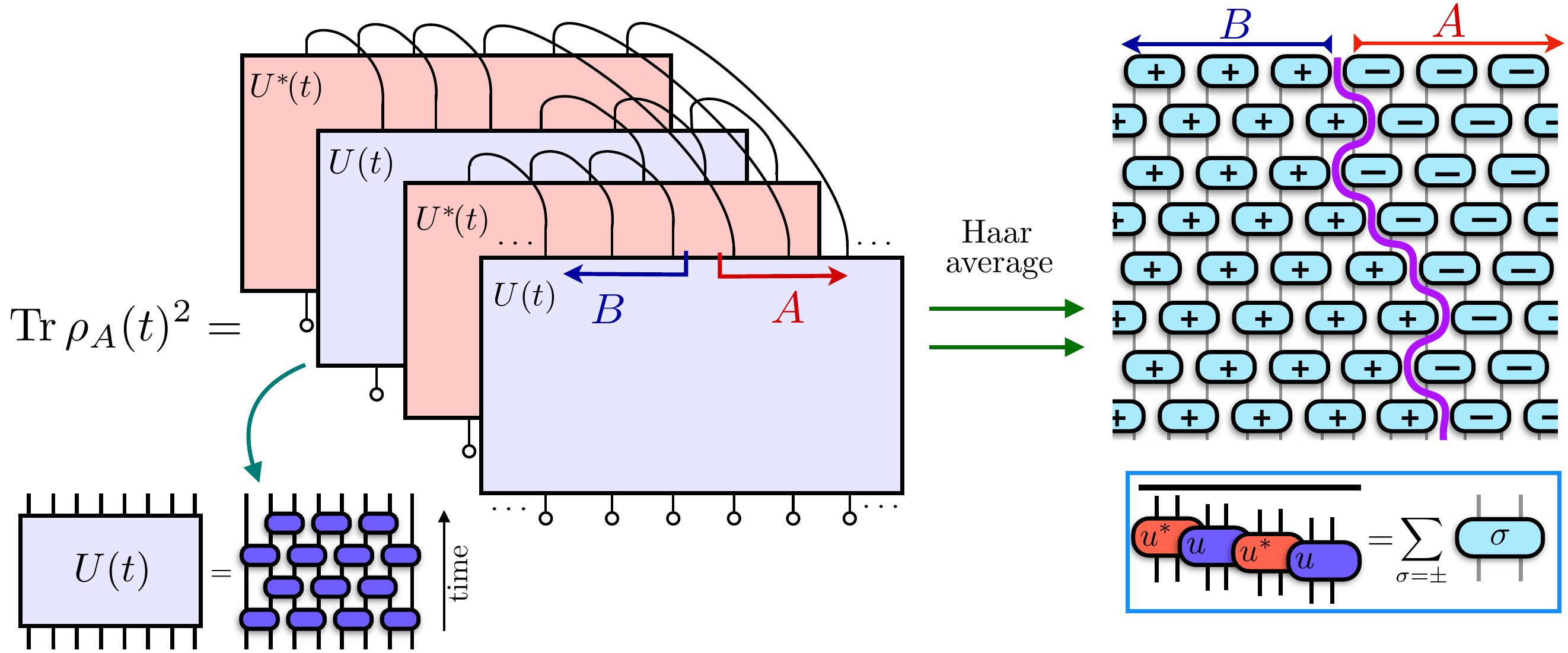} 
\caption{Performing an average over the Haar measure, for each local unitary gate in the brickwork quantum circuit,
transforms $\mathrm{Tr}\rho_{A}(t)^{2}$ into the partition function of a type of two-dimensional classical Ising model, 
with a single Ising spin for each two-site quantum gate.}
\label{fig:haar_avg}
\end{figure}

The R\'enyi entropies are nonlinear functions of the density matrix: e.g. the second R\'enyi entropy $S_{2A}$ is given by (Sec.~\ref{sec:buildingblocks}):
\ba\label{eq:Renyitwo}
\exp ( - S_{2A}(t) ) & =   \Tr_A \rho_A(t)^2,
&
\text{with }\rho_A(t)& = \Tr_B \rho(t).
\end{align}
Again, consider a  quench from a product state. As a tensor network, the evolving state $\ket{\psi}_t$ is expressed using a single copy of the circuit (with product states attached to the bonds at the bottom, and the bonds at the top being the wavefunction's spin indices).
The density matrix ${\rho(t) = \ket{\psi}_t \bra{\psi}_t}$ involves the bra and the ket, so can be represented as a \textit{two-layer} circuit, where the ``bra'' layer is complex conjugated. 
The quantity $\Tr \rho_A(t)^2$, known as the purity of region $A$, is thus a \textit{four-layer} circuit, with two ``backward'' (complex conjugated) and two ``forward'' layers. As a result of the traces in (\ref{eq:Renyitwo}), these layers are sewn together at the final time boundary in the manner shown in Fig. \ref{fig:haar_avg}.  This is a discrete version of a path integral with two backward and two forward trajectories. Let us label the forward layers $1$ and $2$ and the backward layers $\bar 1$ and $\bar 2$.

This structure describes time evolution going on in parallel in four ``universes'', one for each copy of the circuit. Correspondingly, there is now in total a $q^4$-dimensional Hilbert space associated with each spatial position, and each physical gate yields a replicated gate ${u\otimes u^* \otimes u \otimes u^*}$ in the multi-layer circuit, where $u^*$ is complex-conjugated in the computational basis.
However if we consider the average over random unitaries, $\overline{ \Tr \rho_A(t)^2}$, there is a remarkable simplification. The average can be performed separately for each local gate~$u$, since these are independently random. (In the following we will be schematic, see \cite{nahum2018operator,zhou_operator_2020} for more careful exposition of the effective lattice model.)

For simplicity,  consider a single-site  rather than a two-site gate (this does not change the basic point).  The Haar average of ${u\otimes u^* \otimes u \otimes u^*}$ is a \textit{projection operator} onto the subspace of the $q^4$-dimensional Hilbert space that is invariant under ${u\otimes u^* \otimes u \otimes u^*}$ for every $u$ \cite{harrow2013church}. This is spanned by only two states, denoted $\kket{+}$ and $\kket{-}$. Schematically,
\be\label{eq:projector}
\overline{u\otimes u^* \otimes u \otimes u^*} = P_+ + P_- = {\sum}_{\sigma=\pm} P_\sigma.
\ee
This decomposition involves non-orthogonal projectors onto $\kket{+}$ and $\kket{-}$.

The states  $\kket{+}$ and $\kket{-}$ have a simple physical interpretation. 
Formally, they are obtained by pairing up the four layers  in one of two possible ways illustrated below, and forming a maximally entangled state for each pair:
\ba\label{eq:plusminusstates}
+:\quad &\contraction[1.5ex]{}{1}{}{\bar 1}
\contraction[1.5ex]{1\bar 1}{2}{}{\bar 2}
1\bar 1 2 \bar 2,
&
-: \quad&\contraction[2ex]{}{1}{\bar 1 2}{\bar 2}
\contraction{1}{\bar 1}{}{2}
1\bar 1 2 \bar 2.
\end{align}
In the language of Feynman trajectories, paired trajectories survive the average because they contribute equal and opposite phases to the path integral.

(As an aside,  the states in Eq.~\ref{eq:plusminusstates}
generalize the even simpler version for just two copies which we denote $\kket{\mathbb{1}}$: 
\ba\label{eq:plusminusstates}
\mathbb{1}:\quad &\contraction[1.5ex]{}{1}{}{\bar 1}
1\bar 1.
\end{align}
In components, $\mathbb{1}_{ab} = \delta_{ab}$. The invariance ${(u\otimes u^*) \kket{\mathbb{1}}=\kket{\mathbb{1}}}$ is easily checked to be a restatement of the unitarity of $u$.)

Eq.~\ref{eq:projector} means that we can trade the unitary average for a \textit{sum} over an Ising-like degree of freedom $\sigma=\pm$. 
Doing this for every gate in the brickwork circuit of Fig. \ref{fig:haar_avg} maps ${\Tr_A \rho_A(t)^2}$ to the partition function of a 2D classical Ising-like model, with one Ising spin for each physical gate.
The ``Boltzmann weight'' of this Ising model, for a given $\sigma$ configuration, is obtained by contracting a tensor network made up of two-site projection operators, which yields \textit{local} interactions for the $\sigma$ spins. These interactions are anisotropic, and encode hard constraints on the spin configuration that ensure observables are consistent with the underlying unitarity of the dynamics.

Generalizations of these pairing of degrees of freedom are important for random unitary circuits \cite{nahum2018operator, zhou2019emergent}, random tensor networks \cite{hayden2016holographic, vasseur2019entanglement}, random Floquet circuits \cite{chan2018solution}, and monitored circuits \cite{jian2020measurement, bao2020theory,nahum2021measurement,li2021entanglement}, and can even be defined in non-random circuits \cite{zhou2020entanglement}. Here, unitarity imposes structure that radically simplifies the resulting Ising model. 
In outline, $\overline{e^{-S_{2A}}}$ becomes a partition function for Ising configurations  with a single directed domain wall, as illustrated in Fig. \ref{fig:haar_avg}.
This domain wall is reminiscent of the minimal cut discussed above, but in addition to an ``energy'' it has a nontrivial ``entropy'' associated with ``thermal'' fluctuations in the effective model. 
{ If ${q\rightarrow \infty}$ then the ``energy'' term dominates, and this reproduces the minimal cut result.} 
For ${q<\infty}$ both ``energy'' and ``entropy''  contribute to a coarse-grained  line tension $\mathcal{E}(v)$ which can be easily computed.\footnote{Note that in the brickwork circuit the shape of the minimal cut is highly degenerate. 
This is why we instead used the randomly structured circuit to discuss the $q=\infty$ limit.
The degeneracy in the brickwork circuit corresponds to  $\mathcal{E}(v)$ becoming a trivial $v$-independent constant (for ${|v|<1}$) in the limit $q\rightarrow \infty$. 
However once $q$ is finite, $\mathcal{E}(v)$ is nontrivial, so the degeneracy is resolved for the coarse-grained entanglement membrane. Dual unitary circuits  are a class of systems where $\mathcal{E}(v)$ is flat even at finite $q$ \cite{bertini2019entanglement,zhou2020entanglement}.}

{

The averaged purity is only a starting point.
First, in the random circuit, we would really like to compute $\overline{S_{2A}}$ rather than  $\overline{e^{-S_{2A}}}$, because 
 the exponential average can be dominated by rare realizations.
Second, we would like to use the structure above as a guide to calculations in more general models, perhaps without randomness.}

In the random case, $\overline{S_{2A}}$ can be obtained by extending the above calculation using the replica trick \cite{zhou2019emergent} which also plays an important role in random tensor networks \cite{vasseur2019entanglement}. 
Here we give another viewpoint \cite{zhou2020entanglement}, which also extends to circuits without any randomness.

Consider a \textit{particular} circuit instance --- e.g. a particular realization of the random unitaries. Then, for a given gate $u$ in the circuit, we may write the tautology
\be\label{eq:projector2}
{u\otimes u^* \otimes u \otimes u^*} = P_+ + P_- + R_\perp^u.
\ee
Comparing to (\ref{eq:projector}), $R_\perp^u$ is a ``remainder'' left over when ${u\otimes u^* \otimes u \otimes u^*}$ is projected to the non-paired subspace.
 Using this representation for every gate, $e^{-S_{2A}}$ is formally equal to the partition function of a generalized Ising model where the spins take values $+, -, \perp$.  (The spin configurations are again constrained as a result of the underlying unitarity.)
  
{ The weights of configuration with only $+$ and $-$ are the same as before, but local clusters of $\perp$ spins have a nontrivial weight that depends on the local unitaries. 
At first sight this looks complicated, but  a simple picture is recovered after coarse-graining.  We still have a domain wall between $+$ and $-$, but it is dressed on \textit{microscopic} scales by insertions of $\perp$ along its length. This is particularly simple at large but finite $q$, when  $\perp$ spins can be shown to be very dilute. They can be explicitly integrated out, giving a renormalization (of order $q^{-2}$) of the local domain wall cost. 

This renormalization has two effects. 
First, it slightly renormalizes the membrane tension $\mathcal{E}(v)$: 
this reflects the difference between averaging ${S_2}$ and averaging its exponential. Second, it means that the local cost of the domain wall varies from place to place in the random circuit 
because $R_\perp^u$ in Eq.~\ref{eq:projector2} depends on the local random gate.
This amounts to a random local potential  for the polymer. While the microscopic picture is rather different from the
minimal-cut-in-a-random-environment in Sec.~\ref{sec:mincut},
we obtain the same universal physics at large lengthscales: $S_{2A}$ maps to the free energy of a directed polymer in a random medium, with its characteristic exponents.\footnote{The minimal cut problem is a ``zero temperature'' problem involving only energy minimization, while the domain wall above is effectively at finite ``temperature''. However temperature is an irrelevant perturbation for the DPRM, so this difference does not change the basic exponents.} 
} 

 We may also consider the case where the circuit is translationally invariant.
Eq.~\ref{eq:projector2} allows a nonrandom system to be treated by a kind of ``perturbation theory'' around the random circuit result, taking into account successively larger clusters of $\perp$  \cite{zhou2020entanglement}. This is a way to derive  the membrane  beyond the random circuit context.

The pairing structure 
in Eqs.~\ref{eq:projector},~\ref{eq:plusminusstates} generalizes to arbitrary number of layers,
and can be used to discuss the entanglement entropies with integer R\'enyi index ${n>1}$  (or higher moments of the purity, as required for the replica trick)
and many other quantities.

In the general case with $N$ layers each of $U(t)$ and $U(t)^*$,  the ``spin'' $\sigma$ labelling different patterns of pairing becomes an element  of the permutation group, $\sigma\in S_N$ \cite{hayden2016holographic, vasseur2019entanglement, zhou2019emergent}. The  effective statistical mechanics problem has a symmetry that is in general \cite{vasseur2019entanglement, zhou2019emergent, zhou2020entanglement,bao2021symmetry}
\be\label{eq:replicasymm}
G_N \equiv (S_N \times S_N) \rtimes \mathbb{Z}_2,
\ee
where the two copies of  $S_N$ (acting as ${\sigma\rightarrow g_F^{\phantom{1}} \sigma g_B^{-1}}$ for ${g_F\in S_N}$, ${g_B\in S_N}$)
arise from symmetry under permutations of the forward and backward layers respectively,
and the $\mathbb{Z}_2$ generator is associated with exchange of forward with backward layers and acts as ${\sigma\rightarrow \sigma^{-1}}$.
Domain walls have an interesting combinatorial structure for $N>2$: { 
each domain wall is labelled by a permutation group element, and domain walls can split and recombine in accordance with the rules for composing} permutations \cite{zhou2019emergent, vasseur2019entanglement, chan2018solution, li2021entanglement}.\footnote{
{ 
The $N$-dependent domain wall structure means that distinct R\'enyi entropies have distinct line tensions $\mathcal{E}_n(v)$ in general. At late times, it also explains the  Page subleading correction to $S_{nA}$ in the case where $A$ contains half the total number $L$ of qubits, 
${S_{nA} = 
\seq (L/2) -(n-1)^{-1} \ln C_n}$, where $C_n$ is the $n$th Catalan number: in the domain wall picture, $\ln C_n$ is an entropic factor from counting the number of energetically equivalent domain wall configurations.}}
We will return to these domain walls in Sec.~\ref{sec:measurementdwsec}.

Let us comment on the special case of the
von Neumann entropy.
In the approach sketched around Eq.~\ref{eq:projector2}, the entanglement membrane has a clear meaning --- 
as a domain wall in a well-defined degree of freedom $\sigma$ ---
for the   R\'enyi entropies $S_n$ with ${n>1}$. 
The von Neumann entropy $S_1$ is more subtle, and in the current approach must be treated via an analytic continuation (e.g. of $S_n$ for $n>1$). 
Since this continuation is difficult, and since $S_1$ has a special status among the R\'enyi entropies,\footnote{For example, the higher R\'enyi entropies can fail to be the most natural measures for entanglement in some settings, such as many models with conservation laws \cite{rakovszky2019sub}, where $S_{n>1}$ grows parametrically slower than $S_1$.} it would be desirable to have a more direct construction of the entanglement membrane  for $S_1$.
Apart from the limit of infinite $q$ mentioned above, direct calculations of the von Neumann entropy have so far been possible only for circuits with special structure such as dual unitarity \cite{bertini2019entanglement} (Sec.~\ref{sec:dualunitary}) or  Clifford 
\cite{nahum2017quantum,von2018operator}
(Sec.~\ref{sec:clifford}).
Exact computations of the von Neumann entropy are possible in a completely different kind of ``large $N$" limit, using holography and the HRT prescription for entanglement, in terms of a surface in a higher-dimensional Anti-de Sitter space \cite{hubeny2007covariant,ryu2006holographic,mezei2017entanglement,liu2014entanglement}. In the appropriate scaling limit these results are consistent with the membrane picture above \cite{mezei2018membrane, mezei2020exploring}.

\subsection{Spreading and decay of correlations}

\label{sec:unitary:correlations}

\begin{figure}[t]
\includegraphics[width=\columnwidth]{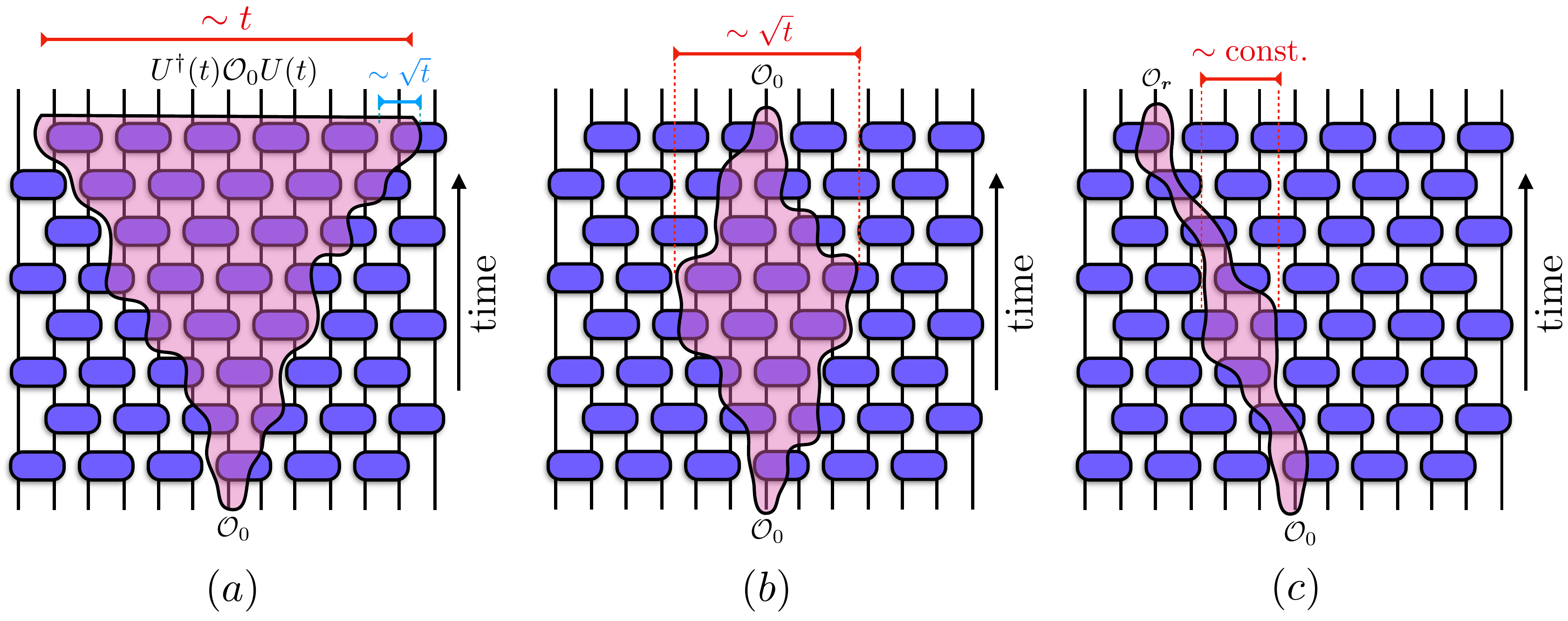} 
\caption{
The typical support of an initially local operator grows ballistically under Heisenberg evolution by a minimally-structured unitary quantum circuit,  while the width of this growing front broadens diffusively in (1+1)-dimensions, as shown in (a). Atypical operator Feynman trajectories contribute to time-ordered correlations of local observables.  Two different classes of atypical trajectories, shown in (b) and (c), can provide dominant contributions to the  mean-squared correlator.}
\label{fig:op_trajectories}
\end{figure}

The evolution of  local observables also exhibits universal structure, for example in the decay of correlations and in the process of ``scrambling'', where information that is initially stored locally becomes delocalized \cite{shenker2014black,shenker2014multiple,maldacena2016bound}. Circuits give an intuitive picture for these processes in terms of Feynman trajectories of  operators~$\mathcal{O}(t)$.

For an $L$-site spin-(1/2) system evolved by the quantum circuit ${U(t) = {U}_{t} {U}_{t-1}\cdots{U}_{1}}$, define the evolution of a local Pauli operator $\mathcal{O}(\boldsymbol{r},0)$  as $ \mathcal{O}(\boldsymbol{r},t) = U(t)\mathcal{O}(\boldsymbol{r},0)U(t)^{\dagger}$ --- for convenience we use the reverse of the usual Heisenberg picture convention here.
Scrambling requires the increase in complexity and spatial extent of $\mathcal{O}(\boldsymbol{r},t)$ with time:  if, say, information is initially stored in the eigenvalue $\sigma$ of the local operator,  ${\mathcal{O}(\boldsymbol{r},0)\ket{\psi} = \sigma\ket{\psi}}$, then at later times this information is stored via the evolved operator,~${\mathcal{O}(\boldsymbol{r},t)\ket{\psi(t)} = \sigma \ket{\psi(t)}}$, which has become delocalized in its support.  

This operator may be expanded as a superposition of ``strings" $\mathcal{S}$, each of which is a product of Pauli matrices at distinct lattice sites, so that \cite{roberts2015localized,ho2017entanglement}
\begin{align}
\mathcal{O}(\boldsymbol{r},t) = \sum_{\mathcal{S}}a_{\mathcal{S}}(t)\mathcal{S}.
\label{eq:op_expansion}
\end{align}
The string operators satisfy $\mathrm{Tr}(\mathcal{S}\mathcal{S}') = \delta_{ \mathcal{S}\mathcal{S}'}$
(the trace $\Tr$ is normalized so ${\Tr \mathbb{1}=1}$)
and the Pauli operator $\mathcal{O}$ obeys  ${\mathrm{Tr}[\mathcal{O}(\boldsymbol{r},t)^{2}] = 1}$.  Therefore, the coefficients satisfy
\begin{align}\label{eq:aanorm}
\sum_{\mathcal{S}}a_{\mathcal{S}}(t)^{2} = 1,
\end{align}
so that $a_{\mathcal{S}}(t)$ may be viewed as the  amplitudes for an evolving, quantum-mechanical ``operator wavefunction" in a $4^{L}$-dimensional Hilbert space 
(each site  can be $\mathbb{1}$ --- i.e. not part of the string --- or $X$, $Y$ or $Z$) \cite{nahum2018operator,von2018operator}. 

The evolution
${\mathcal{O}(\boldsymbol{r},t+1)=
{U}_{t+1}\mathcal{O}(\boldsymbol{r},t){U}_{t+1}^{\dagger}}$
is a linear, unitary evolution of this operator wavefunction, 
${a_{\mathcal{S}}(t+1) = \sum_{\mathcal{S}'}W_{\mathcal{S}\mathcal{S}'}(t)a_{\mathcal{S}'}(t)}$, with the matrix ${W_{\mathcal{S}'\mathcal{S}}(t) \equiv \mathrm{Tr}({U}_{t+1}\,\mathcal{S}\,{U}_{t+1}^{\dagger}\mathcal{S}')}$. 
This operator dynamics simplifies for the circuit composed of Haar-random unitary gates $u_{x,t}$ (Sec.~\ref{sec:circuitdefns}).
In that case, taking the Haar average over the gates $u_{x,t}$ gives 
\begin{align}
\overline{W_{\mathcal{R}\mathcal{R}'}(t)\,\,W_{\mathcal{S}\mathcal{S}'}(t)} = \delta_{\mathcal{R},\mathcal{S}}\,\delta_{\mathcal{R}',\mathcal{S}'}\,T_{\mathcal{S}\mathcal{S}'}(t).
\end{align}
The transfer matrix $T_{\mathcal{S}\mathcal{S}'}(t)$ defines a Markov process on the space of basis operators,
with local updates of the strings and 
string probabilities $\overline{a_S^2}$  \cite{oliveira_generic_2007,dahlsten_emergence_2007,znidaric_exact_2008,harrow_random_2009,nahum2018operator,von2018operator}:
\begin{align}\label{eq:aTa}
\overline{a_{\mathcal{S}}(t+1)^{2}} = \sum_{\mathcal{S}'}T_{\mathcal{S}\mathcal{S}'}(t)
\overline{a_{\mathcal{S}'}(t)^{2}}.
\end{align}
For the initially local operator, this process can be mapped to a classical cluster growth process with an effective stochastic dynamics for the boundaries of the operator  \cite{nahum2018operator, von2018operator}. The initial string is  supported on  one site, e.g. ${\mathcal{O}(\boldsymbol{r},0)=Z_{\boldsymbol{r}}}$. This site forms the seed for a  cluster of Pauli operators that grows with time (in order to exploit the entropy of the cluster in the classical model). The boundaries undergo their own effective stochastic dynamics, which in 1+1D is just biased diffusion.

In (1+1)-dimensions, the two boundaries of the cluster move outwards ballistically with \emph{butterfly velocities}, ${\pm v_B}$; in the limit that the on-site Hilbert space dimension ${q\rightarrow\infty}$, $v_{B}$ approaches the maximum (lightcone) speed ${v=1}$ of the quantum circuit,   
while ${v_B < 1}$ when $q$ is finite (e.g. for spin-(1/2) degrees of freedom).
The growth of an initially local operator can be 
measured with the ``out-of-time-order'' correlation (OTOC) function  $-\f{1}{2}\overline{\mathrm{Tr}([\mathcal{O}(\boldsymbol{r},t),\mathcal{O}(0,0)]^{2})}$ which saturates to $1$ for $|r|\lesssim v_B t$ and is exponentially small for ${|r|-v_B t\gg 1}$.  The OTOC has been used to diagnose the scrambling of locally-accessible quantum information \cite{shenker2014black,shenker2014multiple,maldacena2016bound}, and its growth in time provides signatures of key properties of quantum many-body systems, including their chaotic \cite{aleiner2016microscopic,stanford2016many,gu2017local,roberts2016lieb}, disordered \cite{swingle2017slow,nahum2018dynamics,huang2017out,chen2017out,chen2016universal,he2017characterizing,fan2017out}, or integrable \cite{dora2017out,fortes2019gauging,mcginley2019slow,yan2019similar} nature. OTOCs have been addressed in a range of experiments \cite{zhu2016measurement,swingle2016measuring,yao2016interferometric,garttner2017measuring,li2017measuring}.

In (1+1)-dimensions, the Markov process on strings simplifies substantially, giving rise to a Markov process for the left and right endpoints of an operator string, which perform biased random walks.  
Their stochastic  fluctuations set the scale on which the plateau in the OTOC is rounded, 
so that the ``edge" of the operator broadens diffusively as $t^{1/2}$ in (1+1)-dimensions.  In $(d+1)$ dimensions with $d>1$, operator growth is governed by a Markov process for a $d$-dimensional cluster.  In the simplest case, the stochastic growth of the edge of this cluster is in the  universality class of  the Kardar-Parisi-Zhang equation for a $(d-1)$-dimensional interface, which grows ballistically and broadens in time as $t^{\beta}$ where $\beta = 1/3$ in $d=2$ spatial dimensions \cite{kardar1986dynamic}.

In (1+1) dimensions, ``typical'' operator trajectories look like Fig. \ref{fig:op_trajectories}a. 
By contrast, time-ordered correlators such as 
${G(\boldsymbol{r},t) \equiv \mathrm{Tr}[\mathcal{O}(\boldsymbol{r},t)\mathcal{O}(0,0)]}$ are determined by atypical operator trajectories, and look quite different \cite{nahum2022real}: instead of growing ballistically, the string is small at the final time Fig. \ref{fig:op_trajectories}b and c.  
The non-vanishing average 
$\overline{G(r,t)^2}$  defines a sum over spacetime paths taken by the endpoints, which start at the origin and end at position $\boldsymbol{r}$ at time t.
The atypical trajectories of the endpoints which are required by these boundary conditions give rise to exponential relaxation, $\overline{G(vt,t)^2}\sim \exp(-r(v)t)$,
as expected for a model without slow modes. Physically, this demonstrates that local degrees of freedom are decohered on an order-1 timescale by the other degrees of freedom, which act as a bath.  

However, there is nontrivial structure in the decay rate $r(v)$ \cite{nahum2022real}. In the 1+1D Haar circuit, the dominant contribution to the correlation function changes from trajectories of the endpoints which are ($i$) bound together when $v > v_{B}$ to ($ii$) unbound when $v < v_{B}$.  This ``unbinding" phase transition may be detected in a non-analyticity of the decay rate $r(v)$ as a function of velocity.  

The fact that correlators are dominated by operator trajectories that are more spatially  compact than the support of operator is generic. It can also be observed in dual unitary circuits \cite{kos2021correlations}, and it implies that numerical evaluation of such correlators using the Heisenberg picture is more efficient than ballistic operator growth would suggest \cite{von2021operator,rakovszky2022dissipation}.

\subsection{Structured unitary circuits}

The preceding sections illustrate how minimally structured random circuits help to uncover universal properties of quantum dynamics that follow from unitarity and locality alone. 
However, many physically relevant models (such as time-independent Hamiltonians) have additional structures that must  enter a universal description of the dynamics, for instance    hydrodynamic modes associated with conserved densities. We now discuss  examples of more structured quantum circuits that nevertheless retain analytic tractability. 

\subsubsection{Circuits with continuous symmetries}
\label{sec:u1}
Random circuits can be enriched with continuous global symmetries, for instance they can be constrained to obey a $U(1)$ conservation law for total charge. 
Refs.~\cite{khemani2018operator, rakovszky2018diffusive} studied such a model and obtained an emergent classical description of slow diffusive modes associated with the conserved densities. These slow modes lead to power-law hydrodynamic late-time tails in both time-ordered and out-of-time-ordered correlators. In addition, \cite{khemani2018operator} described the interplay between the slow modes and the other (fast) degrees of freedom to furnish an explicit microscopic derivation of how (reversible) unitary dynamics can give rise to diffusive (dissipative) hydrodynamics. This  addresses a long-standing question in quantum statistical mechanics, namely of reconciling hydrodynamics --- which is a  dissipative classical theory for a few coarse-grained degrees of freedom --- with linear, unitary quantum time evolution on an exponentially large Hilbert space.    

Quantum circuits, by definition, are discrete-time evolutions and hence cannot conserve energy which requires continuous time translation symmetry. Instead, the system can be made to conserve a total charge, for instance total
$S^z_{\rm tot}=\sum_{i} Z_i$.
Each gate conserves  $S^z$ locally, so has a block-diagonal structure, with each block 
(acting within a particular charge sector) 
being an independent Haar-random unitary.
The conservation law induces  non-linearities in the equations for the weights $\overline{a_S^{2}}$
(the analog of \eqref{eq:aTa})
but analytical tractability is  retained in a type of large-$q$ limit, in which each site hosts both the qubit whose $z$ component is conserved 
and an unconstrained large-$q$ degree of freedom \cite{khemani2018operator}.

A local charge density operator at the origin, $Z_0(t) = \sum_\mathcal{S} a_{\mathcal{S}}(t) \mathcal{S}$,  can again be expanded in a basis of Pauli operators. 
In addition to the normalization constraint \eqref{eq:aanorm} on the $4^L$ \emph{weights} $a_{\mathcal{S}}(t)^2$, 
the conservation of $S^z_{\rm tot}$  requires normalization of the $L$ \emph{amplitudes} of the local conserved densities, $Z_i$, in the operator expansion of $Z_0(t)$: 
\begin{align}
      \sum_i a_{Z_i}(t) = 1
\label{eq:norm_conserved}
\end{align}
which follows from 
$\mbox{Tr}[Z_0(t) S^z_{\rm tot}] =  \sum_{i} a_{Z_i}(t) =  \mbox{Tr}[Z_0 S^z_{\rm tot}] = 1$.
After Haar-averaging, the effect of a gate acting on sites $x$, $x+1$ is to spread charge uniformly between these sites:\footnote{Note that the amplitudes of conserved densities survive Haar averaging in a $U(1)$ symmetric system, while only the weights survive averaging in a Haar random circuit with no symmetries.}
\begin{align}
    \overline{a_{Z_x}(t+1)} = \overline{a_{Z_{x+1}}(t+1)} = \frac{\overline{a_{Z_x}(t)} + \overline{a_{Z_{x+1}}(t)}}{2}, 
\label{eq:conserved_diffusion}
\end{align}
Upon coarse-graining, 
Eq. (\eqref{eq:conserved_diffusion}) 
becomes a 
diffusion equation 
 for the conserved quantity $a_{Z_i}$, with diffusion constant $D=1/2$.
This implies that the relaxation
of the conserved density is diffusive, as expected:
$$\mbox{Tr}[Z_0(t) Z_x] = a_{Z_x}(t) = \frac{1}{\sqrt{2\pi t}}e^{-\frac{x^2}{2t}}.$$

\begin{figure}[t]
\centering
\includegraphics[width=.9\columnwidth]{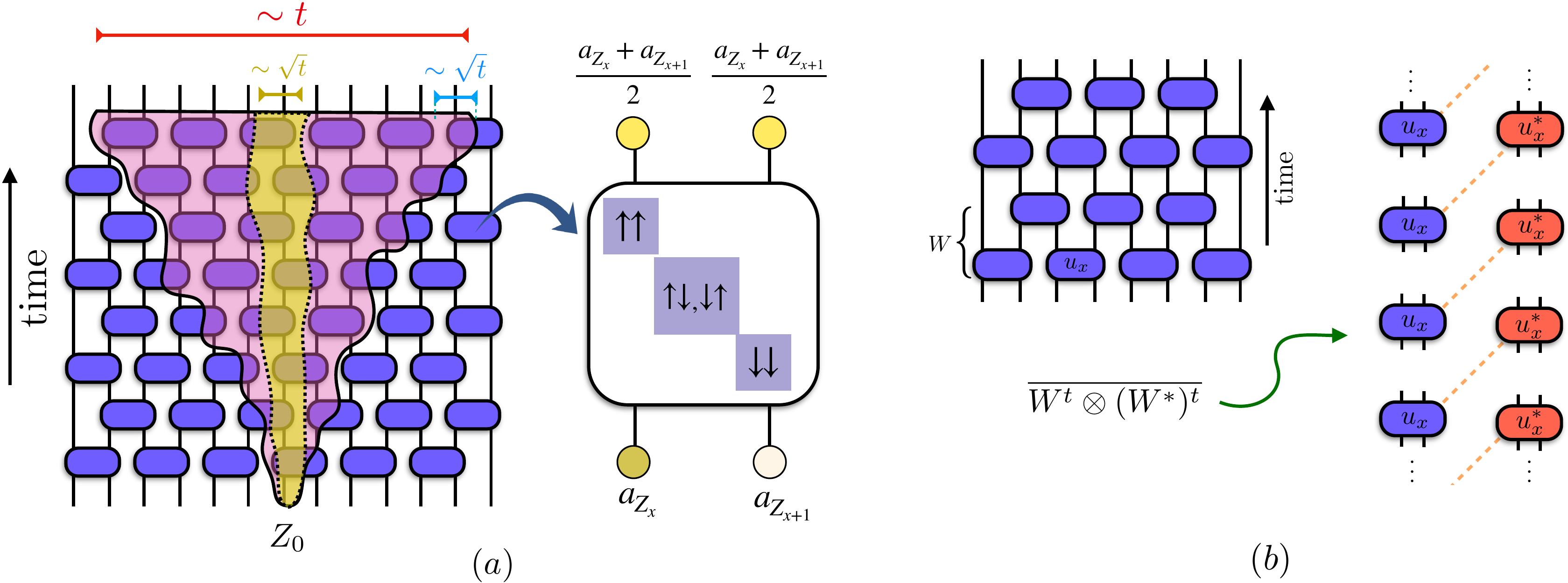}
\caption{The spreading of the local charge density $Z_{0}(t)$ by a quantum circuit which conserves $S^{z}_{\mathrm{tot}} \equiv \sum_{i}Z_{i}$ is shown schematically in $(a)$.  As described in Sec. \ref{sec:u1}, the region of support of this operator grows ballistically, while retaining a weight on a diffusive lump of conserved charges near the origin. A Floquet circuit composed of Haar-random unitary gates is shown in  $(b)$. Averaging over this ensemble in the calculation of the spectral form factor leads to a pairing degree of freedom for each unitary gate in a Floquet period, which simplifies in the limit of large local Hilbert space dimension $q\rightarrow\infty$, as described in Sec. \ref{sec:floquet}.  One of the possible pairings in this limit is indicated by the dashed lines in $(b)$.}
\label{fig:symmetries_floquet}
\end{figure}

The total operator weight ${\sum_{\mathcal{S}} a_{\mathcal{S}}^2 = w^\text{c} + w^\text{nc}}$
may be decomposed into a part $w^\text{c}$ supported on the $\mathrm{U}(1)$ conserved densities $Z_x$, and a part $w^\text{nc}$ supported on more general strings, with ${w^\text{c}+ w^\text{nc}=1}$.
While the sum of conserved \textit{amplitudes} is constant, the weight $w^c$ on the densities decreases as a power law: $\overline{w^\text{c}(t)}=\frac{1}{2\sqrt{\pi t}}$.
This is due to local conversion of conserved to non-conserved operators, at a rate  proportional to the square of the local conserved current.
The emitted nonconserved parts then grow  at the butterfly speed, as before.
This  process can be described via  coupled hydrodynamic equations \cite{khemani2018operator}.

The spreading operator has a characteristic shape that can be detected with the OTOC. The front remains ballistic, but the slow conversion of conserved to nonconserved operators leads to a diffusing lump of local conserved charges near the origin, as shown schematically in Fig. \ref{fig:symmetries_floquet}a,  and also to  power-law tails behind the front: these are due to ``lagging fronts" of nonconserved operators that are emitted at later times during the operator's evolution.

Physically, the emission of long strings allows 
physical observers see a decay of correlations and an increase in observable entropy even though, by unitarity, the  von Neumann
entropy of the full system remains unchanged. 
This is really a decay of correlations detectable with simple operators, due to the  ``hiding'' of correlations in highly nonlocal operators.

The case of $\mathrm{U}(1)$ symmetric diffusive circuits has been generalized to more exotic variants. For instance, higher dimensional circuits with ``subsystem symmetries" acting on lower-dimensional submanifolds can lead to anomalous subdiffusive dynamics~\cite{Iaconis_2019}.  A separate line of work studied ``fractonic" circuits which conserve both a $\mathrm{U}(1)$ charge, $Q= \sum_x q_x$ and its dipole moment, $P=\sum_x x q_x$ \cite{pai2019localization,pai2019erratum,khemani2020localization,sala2020ergodicity}. This was found to lead to a novel form of ergodicity breaking via the ``shattering" of Hilbert space into exponentially many dynamically decoupled sectors, with no dynamical path even between states with the same charge and dipole quantum numbers \cite{khemani2020localization,sala2020ergodicity} --- a surprising result in light of the conventional belief that ergodicity breaking requires extensively many conservation laws, as in integrable systems.

Before leaving this section, we comment briefly on implications of results derived for noisy (random in time) quantum circuits for deterministic dynamics. The broad features discussed here, for instance regarding the behavior of OTOCs and entanglement in systems with and without conservation laws, have numerically been verified to also hold more generally for deterministic dynamics generated by thermalizing time-independent Hamiltonians or time-periodic Floquet circuits. 
(A caveat is that in the presence of conservation laws, the higher R\'enyi entropies can be anomalously affected by rare events in which the local charge is highly depleted, so that they grow and spread more slowly than the von Neumann entropy \cite{rakovszky2019entanglement,huang2019dynamics,
zhou2020diffusive,
feldmeier2020anomalous,gromov2020fracton}.)
One heuristic picture is that thermalizing systems can act as their own bath and generate a noisy
environment for subsystems \cite{lerose2021influence}.
Alternately, as discussed in Sec.~\ref{sec:entanglement:pairings}, the effective degrees of freedom that can be identified microscopically in the random circuit also emerge after coarse-graining in more general models.

\subsubsection{Floquet Circuits}
\label{sec:floquet}
An important class of structured circuits possesses \emph{discrete time-translation symmetry}. If in  the brickwork circuit (Sec.~\ref{sec:circuitdefns}) 
the first two layers of gates are repeated periodically, then 
${U(2t;0) = W^t}$ 
where the evolution operator for one period is  ${W = U_2U_1}$.
Any initial state can be expanded in the eigenbasis of $W$, so that the spectrum of $W$ plays a similar role to that of a time-independent Hamiltonian for continuous time-evolution.

A large class of ``chaotic'' many-body Floquet systems without conservation laws thermalize to infinite temperature, like the fully random circuits discussed in Sec.~\ref{sec:circuitdefns}. 
While we will focus here on chaotic Floquet dynamics, it should be noted that random Floquet circuits can also exhibit many-body localization~\cite{chandran2015semiclassical,Lazarides14, ponte2015many, chan2018spectral,garratt2021many}.
In particular, Ref.~\cite{chandran2015semiclassical} emphasized the quantum circuit language, constructing a  random  Floquet circuit of Clifford gates (Sec.~\ref{sec:clifford}) which showed a solvable localization transition, related to classical percolation of spatial puddles.
For localized Floquet dynamics, the system can avoid thermalizing to infinite temperature, allowing for novel dynamical phases such as the Floquet time crystal phase, which spontaneously breaks discrete time-translation symmetry \cite{khemani2016phase,else2016floquet}, and in 2D the anomalous Floquet insulator, with protected chiral edge modes  \cite{nathan2019anomalous,titum2016anomalous, HarperReview2020}.

We turn now to a headline result in the study of chaotic Floquet quantum circuits, which is 
a demonstration of the random matrix spectral statistics that are often taken as a practical definition of quantum chaos. 
There is a plethora of numerical evidence that spectral correlations in chaotic many-body systems are universally described by random matrix theory (RMT)  \cite{bohigas1984characterization, DAlessio-Rigol2016_review},
but an explicit derivation away from a semi-classical single-particle limit has been a longstanding open problem.
This has been achieved in both Haar random brickwork Floquet circuits in the limit of large local Hilbert space dimension $q$ (below) \cite{chan2018solution,chan2018spectral} --- see also Ref.~\cite{kos2018many} in a slightly different setting --- and in certain ``dual unitary" Floquet circuits with no large parameters \cite{Bertini_2018, bertini2021random,flack2020statistics} (next subsection).
Themes discussed in previous sections, including the pairing of Feynman trajectories and the mapping of circuits to effective classical statistical models, feature centrally in  both results. 

Within RMT, a standard probe of correlations
between the eigenvalues  $\{e^{i\theta_n}\}$ of  $W$
is the spectral form factor (SFF),  which is defined as 
\begin{align}
    K(t) = 
    \left|\mbox{Tr}\,W^t\right|^2 = \sum_{m,n}e^{i (\theta_n - \theta_m)t}.
\end{align}
If $W$ is a Haar-random unitary acting on the full Hilbert space of $L$ qudits, each with Hilbert space dimension $q$, then the SFF averaged over this ensemble is ${\overline{K_{\rm Haar}(t)} = t}$
for  ${1 \leq  t \leq q^L}$. This linear growth (``ramp'') in the SFF characterizes the level repulsion between pairs of eigenvalues across the entire eigenspectrum within RMT \cite{mehta2004random}.

Turning to the case where $W$ describes a quantum circuit composed of local unitary gates,  
$K(t)$ becomes a two-layer circuit with periodic boundary conditions in the time direction, and can be written as a double sum over periodic Feynman trajectories, one in each layer.  Terms in which the forward and backward trajectories are equal \textit{up to a time translation} are special, in that their phases cancel and they give a positive contribution to the sum. Heuristically, the behavior $\overline{K(t)}=t$ 
arises from the fact that there are $t$ ways in which the forward trajectory can be translated with respect to the backward trajectory. Pairings of  trajectories which are not related by a time-translation can contribute any phase to the SFF, and within a given Floquet dynamics, the large number of such pairings can combine to provide a contribution to the SFF which is of the same order as the ramp.  As a result, the SFF is not a self-averaging quantity \cite{prange1997spectral}.  Averaging the SFF over a time-window, or over an appropriate ensemble of Floquet dynamics, is required in order to eliminate these contributions. 

This understanding can be made concrete in random Floquet circuits in (1+1)-dimensions \cite{chan2018solution,chan2018spectral,garratt2021local}.  
Averaging over the Haar-random gates in this setting gives a sum over diagrams involving pairings between the layers. 
In general these may be extremely complicated, but in the limit ${q\rightarrow \infty}$ there is a great simplification. 
The diagrams can be represented as pairings between  between the gates of the forward and backward layer, 
and only $t$ diagrams survive \cite{chan2018solution,chan2018spectral}, which are labeled by a time-shift ${s\in\{0,\ldots,t-1\}}$:
 a given unitary  $u_{x,\tau}$ in the forward layer is paired with $u^*_{x,\tau+s}$ in the backward layer. An example of such a pairing is shown in Fig. \ref{fig:symmetries_floquet}b. For a fixed system size $L$ and at sufficiently long times, the pairing degrees of freedom for each Haar-random unitary gate in $W$ are forced to be identical \cite{chan2018solution,garratt2021local,chan2021many}. Summation over the $t$ possible choices recovers the RMT behavior $\overline{K(t)} = t$.  
 
  A modified  $q\rightarrow\infty$ model contains nontrivial finite-time corrections to the SFF  \cite{chan2018spectral}. In this model the SFF maps to  the partition sum for a one-dimensional, classical $t$-state Potts model, where the Potts degrees of freedom label a spatially-varying, local choice of pairing \cite{chan2018spectral}.\footnote{ Potts symmetry is an artefact of $q\rightarrow\infty$. In general the symmetry is that of a clock model.}
 A disordered regime of the Potts model
  (with domain walls between Potts states)
 arises at sufficiently short times $t<t_{\mathrm{Th}}$,  
 where $t_{\mathrm{Th}}\sim \log L$ was termed a  ``Thouless time''. For $t\gg t_{\mathrm{Th}}$ the Potts model is ordered, and reproduces the ramp behavior expected from RMT.

\begin{figure}[t]
\centering
\includegraphics[width=.84\columnwidth]{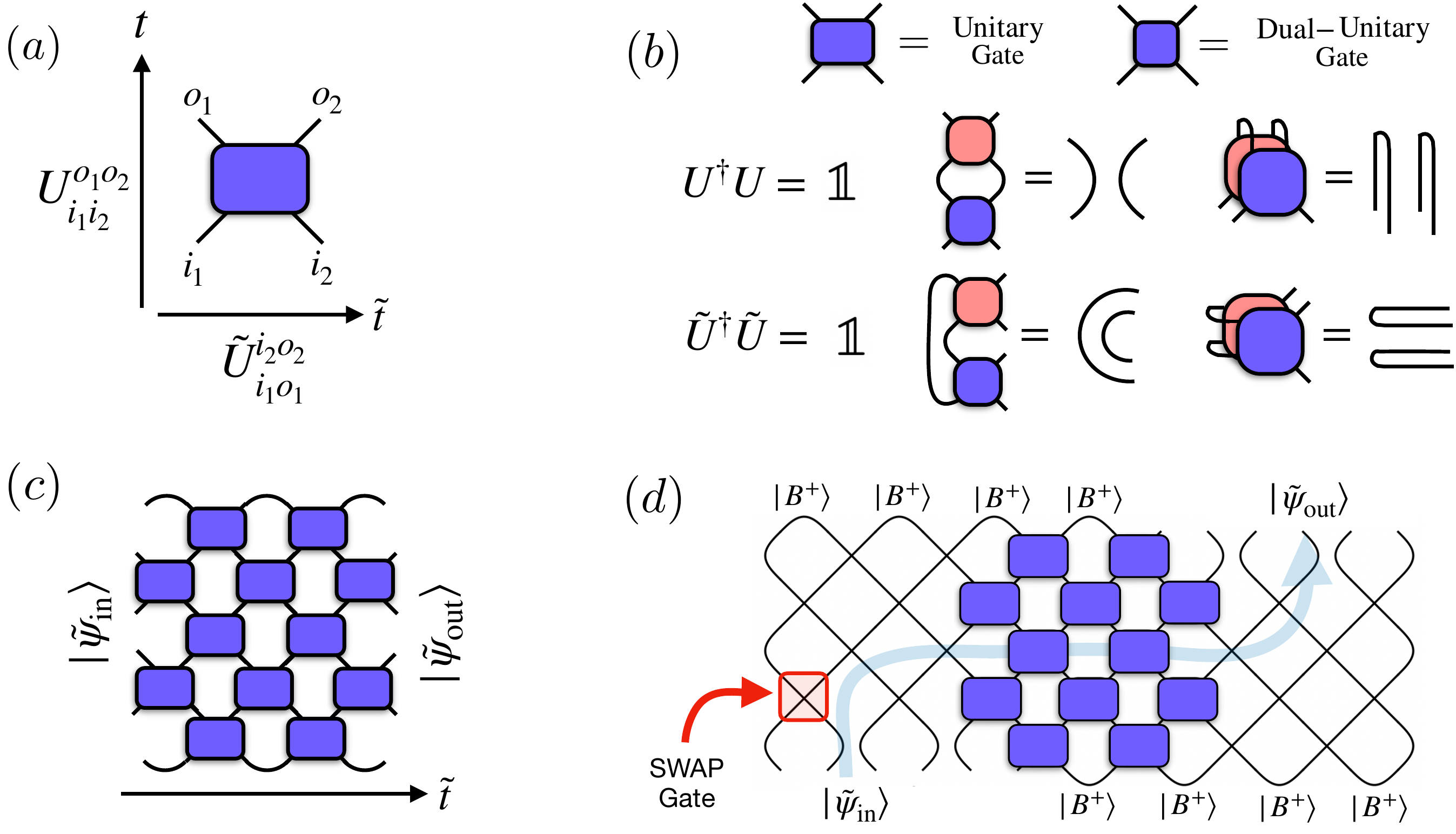}
\caption{A generic unitary gate acting in the time direction $(a)$ can be viewed as a non-unitary transformation acting in the spatial direction. A restricted class of ``dual-unitary" gates act unitarily in both the space and time directions, as shown schematically in $(b)$ (see Sec. \ref{sec:dualunitary}). Unitary circuits give rise to a non-unitary evolution in the spatial direction $(c)$, which can be thought of as unitary evolution interspersed with post-selected measurement outcomes.  This gives rise to a protocol outline in $(d)$ for overcoming the post-selection problem inherent in studying the entanglement properties of monitored pure-states, as discussed in Sec. \ref{sec:monitored} and Sec. \ref{sec:flipped_monitored}.}
\label{fig:dual_unitary}
\end{figure}

\subsubsection{Dual-Unitary Circuits}
\label{sec:dualunitary}

Dual-unitary circuits \cite{Bertini_2018, bertini2019exact, Gopalakrishnan_du} are built up of certain classes of special two-site unitary gates that look unitary in both ``spacelike" and ``timelike" directions. A two-qudit gate
$U_{i_1i_2}^{o_1o_2}$ (unitarily mapping two inputs $i_{1,2}$ to two outputs $o_{1,2}$ according to ``arrow of time'' $t$), could alternatively be viewed ``sideways'', according to a rotated ``arrow of time'' $\tilde{t}$, mapping input qubit states $i_1,o_1$ to output states $i_2,o_2$ as in Fig. \ref{fig:dual_unitary}a.
The resulting map, $\tilde{U}_{i_1o_1}^{i_2o_2}$--- called the ``spacetime-dual'' or ``spacetime-flipped'' version of $U$---is in general \emph{not} unitary (a fact that will be exploited in Section~\ref{sec:flipped_monitored}). However, circuits built from special class of dual-unitary gates, for which the spacetime dual \emph{is} unitary (Fig. \ref{fig:dual_unitary}b), have come to exemplify a form of ``maximal chaos". Various properties such as emergence of RMT spectral correlations in the SFF \cite{bertini2021random}; the dynamics of state and operator entanglement \cite{reid2021entanglement,Gopalakrishnan_du, bertini2020operator,piroli2020exact,zhou2022maximal,ho2022exact};  unequal time correlation functions, both time-ordered and out-of-time-ordered \cite{bertini2019exact,kos2021correlations}; and the eigenstate thermalization hypothesis \cite{fritzsch2021eigenstate,claeys2021ergodic} etc. can be derived in such systems without any approximation and with no small parameters. 

A remarkable property of these circuits is their double causal structure.
Unitarity and the locality of the gates forbid correlations between points displaced by a spacetime vector $(x,t)$ if $|x|>|t|$, i.e. if the two points lie outside each other's (past or future) light cone. 
By the same token, however,   dual-unitarity  rules out correlations if $|t|>|x|$, so that nonzero correlations are only possible on the ray $|x|=|t|$.
This case with two ``arrows of time" can be expanded to cases with more arrows of time~\cite{Jonay_2021}, which may allow for more exotic analytically tractable dynamics.\footnote{We note that tensors with any even number of legs that are unitary under any bipartitioning of the legs  (which includes dual-unitary gates as a subset) were introduced in Ref.~\cite{harris2018calderbank} as ``block-perfect tensors'' in the context of holographic quantum error correcting codes, in Ref.~\cite{berger2018perfect} as ``perfect tangles'' for modular tensor categories, and in in Ref.~\cite{doroudiani2020planar} as ``planar maximally entangled states''.
See Ref.~\cite{borsi2022construction} for discussion of construction of dual unitary gates.}

As discussed in Sec.~\ref{sec:entanglement:pairings}, unitarity implies a simple action of the multi-layer circuit on ``paired'' states. 
Dual-unitarity implies that this also holds for the flipped evolution, 
and this property underlies much of the tractability of dual-unitary circuits. 
Writing the  SFF in terms of a transfer matrix for spatial propagation, its
unitarity  implies the existence of $t$ unit eigenvalues representing paired states (in this case the pairing is between a single $U$ layer and a single $U^*$ layer, with an arbitrary  time-shift in the pairing analogous to that discussed above).
After averaging over an ensemble of chaotic dual-unitary circuits, these eigenvalues dominate in the limit $L\rightarrow \infty$, meaning that in these models there is no long Thouless time and the RMT result holds for finite $t$ as $L\rightarrow \infty$ \cite{bertini2019exact}.  
The spatial transfer matrix has been investigated in more general settings in Refs.~\cite{banuls2009matrix,
akila2016particle,
piroli2017integrable,
garratt2021local,garratt2021many, lerose2021influence}.

\subsubsection{Classically Simulable Circuits}
\label{sec:clifford}

We now consider special classes of quantum circuits which are efficiently simulable on classical computers, with simulation times scaling only {polynomially}, rather than exponentially, with the number of qubits. 

The first: \textit{Clifford circuits}, or ``stabilizer circuits'', play an important role
in quantum information theory. 
Clifford circuits 
are composed of a restricted set of quantum gates with the property that a Pauli string is mapped
under Heisenberg evolution
onto another Pauli string, rather than a superposition of such strings. 
As a result, only a {single} coefficient $a_S(t)$ is non-zero at each time in the expansion of the time-evolved Pauli operator in Eq.~\ref{eq:op_expansion}, and the \emph{operator} entanglement entropy remains zero for all times\footnote{The operator entanglement entropy is computed by treating operators as states in a doubled Hilbert space, for instance, $O = \sum_{ab} O_{ab} |a\rangle\langle b | \rightarrow \sum_{ab} O_{ab} |a\rangle \otimes |b\rangle$.}.
Clifford gates form a discrete subgroup of the unitary group on qudits 
with prime dimension $q\ge 2$.  Here, we restrict our attention to Clifford circuits on qubits ($q=2$).

The Clifford property gives an efficient way of storing a class of states \cite{gottesman1996class,Gottesman_Knill,aaronson2004improved}. 
A `stabilizer state' $\ket{\psi_S}$ on $L$ qubits is defined by $L$ independent and commuting Pauli string operators, the `stabilizers' $\{\mathcal{S}_1, \mathcal{S}_2\cdots \mathcal{S}_L\}$, under which the state is invariant: ${\mathcal{S}_i|\psi_S\rangle = |\psi_S\rangle}$.\footnote{A simple example is a polarized state $\ket{\uparrow, \ldots, \uparrow}$, for which we can take $\mathcal{S}_i = Z_i$.}
At any later time, $\ket{\psi_S(t)}$ is specified by the time-evolved stabilizers 
${\mathcal{S}_i(t)=U(t)\mathcal{S}_i U(t)^\dag}$.
Each $\mathcal{S}_{i}(t)$ operator is a product of Pauli operators on at most $L$ sites.  
Specifying the wavefunction through the stabilizers $\{\mathcal{S}_{i}(t)\}$ is efficient, requiring only $O(L^{2})$ bits of information to store, even if the wavefunction is highly entangled; this is in contrast to the $O(\exp(L))$ cost of storing a generic, highly-entangled state.\footnote{The evolution of generic --- non-stabilizer --- states cannot be efficiently simulated even when the dynamics is Clifford.}
It also allows expectation values and projective measurements of Pauli string operators to be efficiently implemented, and entanglement entropies to be efficiently computed (the R\'enyi entropies $S_n$ are $n$-independent for stabilizer states).

Random unitary Clifford evolution of an initial  product stabilizer state leads  to ballistic growth of entanglement entropy,
as in the Haar circuit. 
In 1+1D this can be understood in terms of the growth of the stabilizer strings $\mathcal{S}_i(t)$ \cite{nahum2017quantum}. 
The choice of stabilizers is not unique  (since if $\mathcal{S}$ and $\mathcal{S}'$ are stabilizers, so is $\mathcal{S}\mathcal{S}'$), and the need to impose a convenient ``gauge" turns the evolution of the stabilizers 
(and their spatial footprints)
into a collective stochastic process, which has analogies to the asymmetric exclusion process for hopping particles on the line.

The time evolution operator for a Clifford circuit belongs to the Clifford group, which is a discrete subgroup of the unitary
group on the full Hilbert space. This group may be generated by a small set of local Clifford gates: the two-site CNOT gate and the single-site Hadamard and Phase gates: 
\begin{equation}
    {\rm CNOT}  = e^{i\frac{\pi}{4} (1-Z_1) (1- X_2)}, \hspace{.25in} 
        H = (X+Z)/{\sqrt{2}}, \hspace{.25in}
P = \sqrt{Z}.    
\end{equation}
{ Any multi-qubit Clifford operation may be written as a product of these gates. } 
Random Clifford circuits can be built by drawing $u_{x,\tau}$ from the set of one and two qubit Clifford gates (or from the set of generators). 
The Clifford gates do not form a universal gate set (i.e. there exist unitary gates which cannot be performed with Clifford operations). However, the Clifford group augmented with { any gate outside of the Clifford group forms a}  universal gate set \cite{nielson2000quantum}; augmenting with a single-site phase shift gate, $T = Z^{1/4}$, is sufficient, for example. 

Despite their non-universal nature, Clifford circuits have proven to be very useful for efficiently simulating certain aspects of quantum dynamics. As an example, even though operator evolution is very special under Clifford dynamics,  \emph{averaging} over the ensemble of uniformly random Clifford circuits can exactly reproduce Haar averages for certain quantities such as OTOCs and right/left endpoint densities of spreading operators \cite{nahum2018operator,von2018operator}. This follows from the fact that the Clifford group forms a unitary 3-design \cite{webb2016clifford,zhu2017multiqubit}, and hence exactly  reproduces averages involving the first three moments of the unitary group. On the other hand,  fluctuations of these quantities within a realization look very different across both classes of circuits. 
There are other systems where Clifford dynamics fails to capture the essential aspects of the problem; for example, Clifford circuits with $U(1)$ symmetry is highly restricted and only shows diffusive spreading for \emph{all} operators, lacking the rich interplay between diffusive conserved densities and ballistic operator growth discussed in Section.~\ref{sec:u1}. 

As an aside, it is worth noting that the Clifford structure relies 
crucially on discreteness of time. In the random circuits we started with, discrete time  also allowed a key simplification, of using the Haar measure for unitaries. 
However random  dynamics can also be formulated in the continuous time limit. An example is the ``Brownian circuit'' \cite{lashkari_towards_2013,shenker2015stringy,zhou_operator_2020,zhou_operator_2019,xu_locality_2019}: this is a Hamiltonian spin chain where the couplings fluctuate like white noise. Operator spreading in  continuous-time noisy models \cite{rowlands_noisy_2018} is qualitatively similar to what we have discussed in Sec.~\ref{sec:unitary:correlations}.

The continuous time limit is also natural for noisy models of \textit{free fermions}.
Gaussianity is a second  important structure that can be imposed on the dynamics and which leads to efficient simulability.

The ``quantum symmetric simple exlusion process'' (QSSEP) \cite{bauer2019equilibrium,bernard2019open,bernard2021entanglement} is a model of complex fermions hopping on the line, with noisy amplitudes, in which many quantities, for example the moments of Green's functions in late-time nonequilibrium states,
can be calculated exactly. 
The simplest averages map to the classical symmetric exclusion process. 
However, higher-order moments diagnose  quantum coherences, 
with a nontrivial combinatorial structure.
It will be interesting to study the crossover between the free fermion limit and the the strongly interacting regime. (In the replica language, this is the reduction of a continuous to a discrete replica symmetry.)

Finally,  \textit{automaton circuits}  are classically simulable circuits  obtained by promoting reversible classical cellular automata to unitary quantum evolutions.
In the computational basis  of (say) $Z$ eigenstates   $\{|n\rangle\}$, the action of such a circuit is $U|n\rangle = e^{i \theta_n}|\pi(n)\rangle$, where $\pi \in S_{2^N}$ is a permutation of the $2^N$ basis states for $N$ qubits. While the time-evolution of computational basis states is classical, these circuits can generate volume law entanglement 
when acting on product states that are not computational basis states. 

Automaton circuits provide a classically-tractable setting in which to observe a range of interesting dynamical phenomena.  First, under evolution by an automaton circuit (for say, spin-(1/2) degrees of freedom), Pauli operators can evolve into sums of products of Pauli operators, unlike in Clifford circuits.  Nevertheless, the operator wavefunction evolves according to the same automaton circuit in a rotated basis \cite{iaconis2019anomalous}.  As a result, operator growth is classically tractable to study; 
out-of-time-ordered correlators in random, local automaton circuits, for example \cite{iaconis2019anomalous}, have been shown to exhibit growth and broadening of the operator fronts.
Automaton circuits 
also provide a tractable setting to study subdiffusive hydrodynamics and kinetically-constrained dynamics \cite{iaconis2019anomalous,feldmeier2020anomalous,iaconis2021multipole,richter2022anomalous,singh2021subdiffusion}, 
a type of measurement transition \cite{iaconis2020measurement},
and integrability \cite{Gopalakrishnan_automata, klobas2021exact_1,klobas2021exact_2,lopez2022integrability}.   Floquet automaton circuits have also been a starting point for constructing fully quantum-mechanical Floquet models with exact, non-thermal (``scarred") eigenstates \cite{iadecola2020nonergodic,rozon2021constructing,gopalakrishnan2018facilitated}.    
A notable example 
is the integrable, Floquet circuit corresponding to the Rule 54 cellular automaton, which
implements a simple structured dynamics involving conserved left/right moving solitons. This model captures many features of  more complicated integrable systems,
but permits  solutions of various quantities including  non-equilibrium
steady-states, operator and entanglement spreading, and generalized hydrodynamics~\cite{Gopalakrishnan_automata, Gopalakrishnan_rule54_otoc,Alba_2019, klobas2021exact_1,klobas2021exact_2}.

\section{Monitored dynamics}
\label{sec:monitored}

As discussed in Sec.~\ref{sec:unitary:entanglement}, a system of initially unentangled qubits, subjected to generic unitary gates acting on pairs of qubits, will rapidly entangle, and in the
long-time steady state will have volume-law entanglement entropy with the maximal entropy density per qubit. 
At extremely long  times, the evolving pure state wavefunction
will be essentially random --- sometimes called a Page state.  But when such a system is monitored with repeated local measurements, this inexorable growth of entanglement will be counteracted.
Indeed, a single-qubit projective measurement  disentangles the measured qubit.  
In this Section we consider the ``hybrid" quantum circuits described in Sec.~\ref{sec:measurementdefns},
in which the brickwork of unitaries is decorated with single-site measurements, placed at each space-time point with probability $p$.
Such a structureless minimal circuit has three types of randomness:  in the two-qubit unitary gates, in the locations of the projective measurements, and (more fundamentally) in the intrinsically random outcomes of the measurements, as dictated by the Born probability.

\subsection{Measurement-induced entanglement transition}

A pure initial state, $|\psi(0) \rangle$, 
evolving under a hybrid circuit, defines a set of (normalized) quantum trajectories, 
labelled by the measurement outcomes $\mathbf{m}$:
\begin{equation} \label{eq:monitored:trajectory}
| \psi_{\vec{m}}(t) \rangle  = K_{\vec{m}} | \psi(0) \rangle / \sqrt{p_{\vec{m}}} .
\end{equation}
The circuit $K_{\bf m}$, consisting of unitaries interleaved with projectors, was defined in Sec.~\ref{sec:measurementdefns}.
The Born probability $p_{\vec{m}} = \langle \psi(0) | K_{\vec{m}}^\dagger K_{\vec{m}} | \psi(0) \rangle$ of a trajectory depends on the state, so that the monitored circuit dynamics is both non-linear and non-unitary.

The idea of the measurement-induced transition \cite{skinner2019measurement,li2018quantum,li2019measurement} is that there is a qualitative change in the nature of the typical  trajectories in Eq.~\ref{eq:monitored:trajectory} as a function of the measurement rate $p$.
When measurements are frequent, the stochastically evolving wavefunction is trapped, by the single-spin projections, within the subspace of area-law states.
(The extreme limit of this case is $p=1$, where in each timestep every qubit is measured simultaneously, giving a product state.)
But when $p$ is reduced below a threshold $p_c$, $\ket{\psi}$ escapes into the volume-law part of Hilbert space.
At the transition itself, the evolving state has a random but scale-invariant entanglement structure.
Numerical evidence for this transition has by now been found in a wide range of microscopic models \cite{skinner2019measurement,li2018quantum,li2019measurement,li2021conformal,zabalo2020critical,zabalo2022operator,gullans2020dynamical,gullans2020scalable,jian2020measurement,szyniszewski2019entanglement,chen2020emergent,turkeshi2021measurement, turkeshi2020measurement}. The efficient simulability of the Clifford circuits allows for particulary precise results there (Sec:~\ref{sec:criticalproperties}). The transition is not specific to one dimension and we can even consider ``all-to-all'' models without spatial locality  (Secs.~\ref{sec:purification},~\ref{sec:criticalproperties}).

The discrete spacetime structure of the circuit lets us access a simple ``classical limit'' of the transition, which gives some intuition for how measurements inhibit the propagation of quantum correlations through spacetime.  
We reach this classical limit either by taking $q\rightarrow\infty$, or by considering the somewhat unphysical ${n\rightarrow 0}$ limit of the  R\'enyi entropy $S_n$.
In these limits the computation of entanglement reduces to the minimal cut (Sec.~\ref{sec:mincut}), and therefore to a ``classical'' geometrical problem \cite{skinner2019measurement}.
This limit is non-generic as far as the critical exponents (and the value of $p_c$) are concerned \cite{skinner2019measurement,zabalo2020critical,potter2021entanglement}, but it captures crude features of the transition and phases. 

Viewing  $K_{\bf m}$ as a tensor network,
a projection operator breaks a bond (it is an operator with trivial rank).
For small enough $p$, enough bonds remain unbroken that the circuit is connected on large scales.
But, computing entanglement with the minimal cut, the broken bonds reduce the cut's line tension.
Therefore the volume-law coefficient ${\seq/\ln q}$ in the steady state is reduced from unity once ${p>0}$. Increasing $p$,  we eventually reach the ``percolation threshold'' where the circuit falls apart into finite disconnected pieces. Beyond this point, the disconnected structure of the tensor network representation of $\ket{\psi}$ immediately implies that $\ket{\psi}$ cannot have long-range entanglement (a point also used in Ref.~\cite{aharonov2000quantum}). It also means that the line tension, setting the volume-law coefficient, has a critical vanishing at $p_c^\text{classical}$. At this transition the entanglement of a subregion scales logarithmically with subregion size, a consequence of scale-invariance.

\begin{figure}[t]
\centering
\includegraphics[width=\columnwidth]{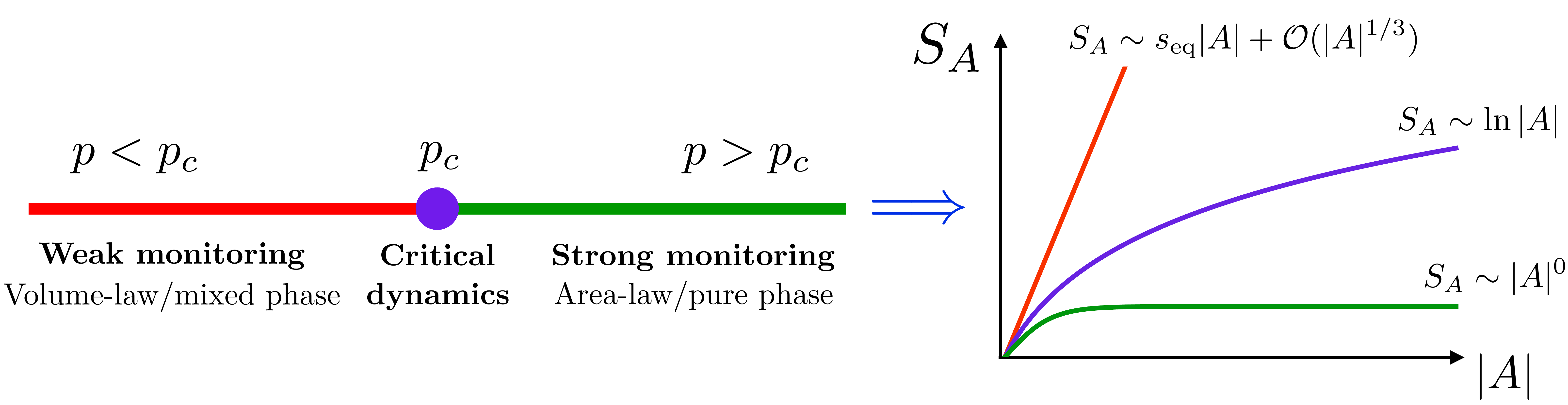}
\caption{Monitored dynamics in (1+1)-dimensions leads to a phase transition in the entanglement properties of the pure-state trajectories as a function of the monitoring rate $p$, as summarized schematically on the left.  The scaling of the von Neumann entanglement entropy in the steady-state of the monitored dynamics with subsystem size, $|A|$, in the two phases and at the ``entanglement phase transition" is shown on the right.}
\label{fig:monitored_evolution}
\end{figure}

Below we will discuss the properties of the phases and the critical point further; a schematic phase diagram along with the scaling of the von Neumann entanglement entropy in these phases and at the critical point in (1+1)-dimensions is summarized in Fig. \ref{fig:monitored_evolution}.
The area-law phase is relatively simple: the states are product states dressed with short-range correlations.
The volume-law states at ${0<p<p_c}$ are  nontrivial, and in particular are \emph{qualitatively different} in their entanglement structure from the Page-random states that are obtained at long time for ${p=0}$ \cite{chan2019unitary,fan2020selforganized}. 

Indeed, neglecting this important difference would lead to the  incorrect conclusion that the volume law phase is unstable for arbitrary small $p$: In a given timestep, a region $A$ suffers an extensive number of measurements ($\propto |A|$), while only of order $|\partial A|$ unitaries act across the boundary of $A$ and so can increase $S_A$. 
If we assumed that each measurement gave an $\mathcal{O}(1)$ decrease in $S_A$ --- which would be true for a Page state --- we would conclude that measurements always win and the volume law state is unstable. 
The failure of this argument is because the volume law states for ${0<p<p_c}$ are dissimilar to Page states, in that the change in $S_A$ from measuring a qubit deep inside $A$ is typically very small. This can be seen in the language of the minimal cut, or more generally in terms of  entanglement domain walls discussed below (Sec.~\ref{sec:measurementdwsec}).

\subsubsection{Coherent trajectories  versus dissipative information loss} 
\label{sec:trajectoryvsmixed}

Under the circuit dynamics, the density matrix ${\rho}(0)$
evolves to ${{\rho}_{\vec{m}} (t) = {K}_{\vec{m}} {\rho}(0) {K}^\dagger_{\vec{m}} / p_{\vec{m}}}$
 with probability ${p_{\vec{m}} = \Tr({K}^\dagger_{\vec{m}}  {K}_{\vec{m}} {\rho} (0) )}$.
It is important to distinguish between the state $\rho_{\vec m}$ --- which obtains when given measurement outcomes ${\bf m}$ have been recorded, and which is a pure state if the initial state is pure ---
from the trajectory-averaged mixed state\footnote{We imagine performing this average with the other parameters of the circuit --- the choices of unitaries and measurement locations --- held fixed.}  ${\overline{\rho}\equiv\sum_{\bf m} p_{\bf m} \rho_{\bf m}}$. 
The latter would be the appropriate description of the system if there was classical uncertainty about ${\vec m}$ --- i.e. if instead of measurements (with a known outcome), 
we had ``decoherence'' due to interaction with an uncontrolled environment.
For the dynamics we consider, $\overline{\rho}$ always tends to the trivial, infinite temperature density matrix:  the rich structure of correlations in individual trajectories is washed out.

The entanglement entropy $S^{\vec m}_A$ of a subregion $A$, conditioned on the measurement outcomes, quantifies its correlations with the outside, and is a probe of the transition, as noted above.
This should be distinguished from the 
``thermodynamic'' entropy of the region, defined as  ${S^\text{th} = - \Tr \bar\rho \ln \bar \rho}$. This will tend to the maximal value even in the area-law phase, because of the additional classical uncertainty (ignorance of the measurement outcomes) implied in using  $\bar\rho$.

Similarly, correlators of the form ${\<\mathcal{O}\mathcal{O}'\>_{\bf m}=\Tr \rho_{\bf m}\mathcal{O}\mathcal{O}'}$ may be nontrivial. 
But trajectory-averaged correlators 
${\overline{\<\mathcal{O}\mathcal{O}'\>_{\bf m}}=\Tr \overline{\rho} \mathcal{O}\mathcal{O}'}$
are trivial at late times because of the triviality of $\bar\rho$.
These statements are consistent  because of cancellation between trajectories in which ${\<\mathcal{O}\mathcal{O}'\>_{\bf m}}$ has opposite signs.
This means that if we want to detect the transition using averaged correlators we need Edwards-Anderson-like (squared) correlators,
$\overline{
\<\mathcal{O}\mathcal{O}'\>_{\bf m}^2
}$.
Like the entanglement measures, these are nonlinear in the density matrix.

To measure these ``squared''-type correlators, pairs of identical quantum trajectories (with the same measurement outcomes ${\vec m}$) will be required, which will generally require post-selection on measurement outcomes. This is exponentially costly, requiring a number of runs exponential in the circuit's space-time volume (but see Sec.~\ref{sec:flipped_monitored} below for an alternate type of monitored dynamics where this postselection barrier is parametrically improved.) .

Above, we viewed the ${\vec m}$ as a record of physical measurements conducted by a hypothetical experimentalist. 
Quantum trajectories also  arise in a quite different setting, 
as formal tool for treating open quantum systems which are in contact with a decohering environment (\textit{not} monitored).
Assuming we do not have access to the quantum state of the environment, such  a system must be described by a mixed state $ \rho_\text{open}(t)$.
In a Markovian approximation for the environment, $\rho_\text{open}(t)$ evolves via a quantum channel, 
$\rho_\text{open}(t)=\Phi_t(\rho_\text{open}(0))$ \cite{nielson2000quantum}.
We can ``unravel'' this open system dynamics by writing $\rho_\text{open}(t)$ as  an average over pure trajectories. 
In a simple case (details omitted), this leads to an ensemble formally equivalent to that discussed above.
Mathematically, this is a rewriting of the channel in terms of Kraus operators  $K_{\bf m}$,
\be
\rho_\text{open}(t) = \sum_{\bf m} 
K_{\bf m} \rho_\text{open}(0) K_{\bf m}^\dag,
\ee
with $\sum_{\bf m} K_{\bf m}^\dag K_{\bf m}= 1$. In this setting, the trajectories are fictitious, but they may be useful for simulating the dynamics \cite{plenio1998quantum,gardiner2004quantum,
daley2014quantum,
bonnes2014superoperators,
garrahan2010thermodynamics}.
In this setting the relevance of 
the measurement transition is as an easy-to-hard transition for classical simulation of various kinds of quantum processes, because the area law states at $p>p_c$ can be efficiently represented using matrix product states, while the highly complex volume-law states at $p<p_c$ are, a priori, exponentially costly  to store and simulate. In the case with true measurements, we can think of the transition into the hard phase as an ``epistemological phase transition'' \cite{nahum2020entanglement} where the wavefunction becomes effectively unknowable, even with access to the measurement outcomes.

\subsubsection{Entanglement domain walls and the stability of the volume law phase}
\label{sec:measurementdwsec}

In (1+1)-dimensions, the universal entanglement properties of the volume law phase can be quantitatively described by the classical statistical mechanics of a domain wall that is equivalent to a directed polymer in a random environment (DPRE) \cite{li2021entanglement}.  The most dramatic outcome of this correspondence is the presence of a universal sub-dominant contribution to the entanglement entropy in the volume law phase, with $S_A = s_0 |A| + b |A|^\beta$ with $\beta = 1/3$.  

The simplest limit of the DPRE is the minimal cut (Fig.~\ref{fig:min_cut}).\footnote{The figure illustrates that in the presence of measurements, the minimal cut is not required to be directed on the microscopic scale. However in the phase where its line tension is nonzero, it is directed on large scales for energetic reasons. The same is true of the domain walls discussed below.}
More generally, the result can  be  understood 
using the effective 2D statistical mechanical model for replicated ``spins" $\sigma$ (defined on the circuit's space-time manifold, with one spin for each physical gate) which label patterns of pairings between the different replicas of the circuit \cite{jian2020measurement,
bao2020theory}. 
We have discussed these spins in the unitary context in Sec.~\ref{sec:entanglement:pairings}, but the non-unitary nature of the monitored dynamics alters the effective  statistical mechanical model. 
One point is that the wavefunction must be explicitly re-normalized after each measurement when the dynamics is not unitary.
More importantly, non-unitarity relaxes local constraints on the $\sigma$ configurations that would follow from unitarity. 
In particular, 
the projective measurements can drive a \textit{disorder transition} for $\sigma$ \cite{jian2020measurement,bao2020theory}.  
We do not give a detailed exposition of the Boltzmann weights for the effective model here, 
but summarize some qualitative features,  starting with the volume-law phase, where   $\sigma$ is ordered.
In this phase, the model is best viewed in terms of domain walls that are forced into the 2D spacetime manifold by the \textit{boundary conditions} required to define the Renyi entropies (recall Fig.~\ref{fig:haar_avg}).

Within the volume-law-entangled phase, the entanglement entropy in a sub-region $A$
can be mapped to the free energy cost, $F_A$, of changing the boundary conditions
at the final time slice of the circuit, within the ordered phase of the statistical mechanics model.  This analytic description, and the associated replica limit which is necessary to obtain the von Neumann entanglement entropy, is quantitatively similar to the replicated description of the DPRE, and is thus conjectured to described the same universal physics \cite{li2021entanglement}. The line tension of the  resulting {\it entanglement domain wall} yields a volume law scaling of the entanglement, $F_A = s_0 |A|$, while the sub-dominant contribution comes from the universal fluctuations of the DPRE free energy with length, with exponent $\beta = 1/3$.  This exponent, believed to be a universal feature of the volume-law phase in the presence of random unitary operators, is found numerically for the hybrid random Clifford circuit.  Universal scaling functions governing the saturation of the entanglement entropy reveal the super-diffusive ``wandering exponent'' $\zeta = 2/3$ of the directed polymer.  Both this exponent and the universal scaling functions extracted numerically from the DPRE are in good agreement with the numerical results obtained from the random hybrid Clifford circuit \cite{li2021entanglement}.

The stability of the volume-law phase against measurements is quantified by considering the reduction $\delta S(x)$ in the entanglement of a region $A$ after performing a local measurement a distance $x \ll |A|$ away from the boundary \cite{fan2020selforganized}.  While $\delta S(x)$ is always an $O(1)$ constant for a Page state, this quantity decays, on average, as a power-law $\langle \delta S(x)\rangle \sim x^{-\Delta}$ in the volume-law phase of the monitored circuit with a consistent exponent $\Delta\approx 1.25$ observed in both the hybrid Clifford dynamics and in the behavior of the corresponding observable in numerical simulations of the DPRE. 
With $\Delta >1$, the total loss of entanglement after a sequence of measurements $\sim \int_{0}^{|A|} \delta S(x) \,dx$ is finite for large $|A|$, and can be recovered with the application of the next layer of unitaries, leading to a stable statistical steady state in the volume law phase.  
We note that the average behavior of $\delta S(x)$ is dominated by rare events and that the scaling of the disorder-averaged free energy of the DPRE provides the prediction that a  \emph{typical} realization of the hybrid dynamics is significantly more robust, with this quantity decaying as a stretched exponential $\delta S(x) \sim \exp[-\mathrm{const}\times x^{\beta}]$.

\subsubsection{Mixed state purification transition}
\label{sec:purification}

The transition between the volume and area law phases can alternatively be understood
as a dynamical ``purification" transition \cite{gullans2020dynamical,bao2020theory,choi2020quantum}.  
For a maximally mixed initial density matrix, 
${\rho}(0) = \mathbb{1}/2^L$, 
measurements at rate $p>p_c$
are able to  purify the state at a system-size-independent rate 
(``pure'' phase), 
while for  $p<p_c$ the purification time diverges exponentially with the system size (``mixed'' phase).   
These two phases correspond to the area and volume law phases of the circuit evolving with an initial pure state. 
In the ``mixed" phase, at times polynomial in the system size there is a residual non-vanishing entropy density,
$s_Q$.\footnote{In the min cut/domain wall picture, this is related to the cost of a horizontal domain wall that separates the initial from the final time \cite{bao2020theory,li2021statistical,nahum2021measurement}, and which measures the ``entanglement'' between the initial and final time boundaries of the non-unitary circuit. This shows that $s_Q$ is also equal to the pure state volume-law coefficient.} Heuristically, $s_Q L$ is the amount of quantum information propagated from the initial to the final time.

This entropy describes a dynamically evolving encoded sub-space which is insensitive to future measurements -- quantum information is effectively protected from single qubit measurements.  
For the hybrid random Clifford circuit,
this corresponds to a ``stabilizer" quantum error correcting code, denoted as $(L,k,d)$,
with $L$ physical qubits, $k$ encoded logical qubits and $d$ the code distance - the size of shortest logical operator.  
Since { $k= s_Q L$}    
is given by the dimension of the encoded sub-space, it grows linearly with $L$,
and the code rate, $k/L \ne 0$, is finite for large $L$.  The entanglement domain wall picture can be used to extract the code distance, which varies as $L^\beta$,
coming from entropic fluctuations of the DPRE.   
The code rate vanishes upon approaching 
the ``pure" phase -- the purification transition thus corresponding
to a transition in the encodability of the quantum code \cite{gullans2020dynamical,li2021statistical,fan2020selforganized,choi2020quantum}.  
While resembling
a quantum error correcting code, since an efficient decoder in the presence
of additional qubit errors is not known, and in fact might not exist,
these codes are not practical.

The purification picture also lends a natural ``local" order parameter for the measurement transition \cite{gullans2020scalable}. The initially maximally mixed state can be viewed as arising from tracing out a set of reference ancilla qubits that are maximally entangled with the system qubits at $t=0$, but do not directly participate in the dynamics thereafter. The purification transition can then be probed by measuring the entanglement entropy of the reference qubits --- actually one reference qubit suffices --- at times $L \ll t \ll e^L$, which is non-zero in the mixed phase but vanishes in the area law phase.

Purification is also a useful way to characterize the measurement transition
for  all-to-all coupled systems of qubits, where any qubit can interact with any other
\cite{gullans2020dynamical,vijay2020measurement,nahum2021measurement}.
For example the  dynamics could consist of  a random sequence of operations, each  being  (with probability $p)$
a measurement of a random  qubit or (with probability $1-p$) a gate applied to a random pair. 
In this setting the volume of a subset $A$  of spins scales the same way as its ``surface area'' (number of potential neighbors outside $A$), so that the volume law coefficient ${S_A/|A|}$ in   a pure state is not a useful diagnostic. 
However, the code rate $s_Q$ above remains a valid order parameter for the transition.

\subsubsection{The critical point}
\label{sec:criticalproperties}

The entanglement transition between the area and volume law phases is a continuous phase transition, and can be explored both numerically and in some instances analytically.

Detailed large-scale numerical simulations are possible on hybrid random Clifford circuits (with two site unitaries drawn randomly from the Clifford group and single site Pauli measurements) and reveal the existence
of a finite critical measurement rate, $p_c \approx 0.17$
in 1+1 dimensions \cite{li2017measuring}, separating a volume law entangled phase for $p<p_c$ from an area law phase when $p>p_c$, as shown in Figure \ref{fig:monitored_evolution}. The critical point has a dynamical exponent $z=1$, and in 1+1 dimensions exhibits a conformal symmetry \cite{li2021conformal,zabalo2022operator}.
Simulations with Haar-random gates or in general Hamiltonian models are restricted to much smaller sizes but are also consistent with a continuous transition with~${z=1}$ \cite{zabalo2020critical,zabalo2022operator}. 

When the qubits are extended to qudits, with dimension $q \rightarrow \infty$, analytic progress is possible, since as  discussed above the phase transition in this limit maps to the problem of a minimal cut through a percolation configuration \cite{skinner2019measurement,potter2021entanglement}.
The critical exponents are then those of the non-unitary conformal field theory for percolation.
However the percolation fixed point is unstable for finite $q$ and the critical properties are in general analytically intractable, though expected still to be conformal.

It is natural to ask for a ``Landau-Ginsburg''--like field theory for the measurement transition.
This is  closely connected to the problem of formulating a field theory for the entanglement transition in a random tensor network (RTN) \cite{vasseur2019entanglement}
(a random tensor network, with ``virtual'' bonds in the bulk and ``physical'' bonds on the boundary, can undergo an entanglement phase transition as a function of the distribution of local tensors) though there is a key difference related to Born's rule.

At first sight we might think that a field theory could be inferred immediately from the  effective lattice magnet, mentioned above,  for the degree of freedom $\sigma\in S_N$.
This formalism is powerful in the ordered/entangling phase, where we can work with domain walls, 
and gives a useful magnetic analogy for the critical point, 
but it is  more challenging to coarse-grain $\sigma$ to arrive at a continuum theory near the critical point.
This is because of the need for\footnote{In Sec.~\ref{sec:entanglement:pairings} we sketched a mapping of $S_2$ to a domain wall free energy that avoided the replica trick. This used the fact that clusters of  ``$\perp$'' spins were of  finite typical size, rather than proliferating in spacetime. 
This approach continues to  hold in the entangled phase in the presence of measurements \cite{nahum2021measurement}, but not at the critical point, where the typical size of $\perp$ clusters diverges.} the replica trick  \cite{vasseur2019entanglement,jian2020measurement, bao2020theory,nahum2021measurement}:
disorder averages must be handled by analytically continuing the lattice magnet to ``unphysical'' values of $N$, where the target space $S_N$ for $\sigma$ collapses to nothing\footnote{For heuristic motivation, recall a simpler use of the replica trick for averages in disordered systems. Let $Z$ be a partition function for a classical field $\phi$, with random couplings. The averaged free energy may be written as ${\overline{\ln Z} = 
\lim_{N\rightarrow 0} (\overline{Z^N}-1)/N}$. On the RHS, $\overline{Z^N}$ is a partition function for $N$ copies of the field, which we can arrange as a vector ${{\mathbf{\phi}}=(\phi_1,\ldots, \phi_N)}$. The RHS requires us to take the limit of vanishing number of components.} (see Ref.~\cite{potter2021entanglement} for a review).

Effective lattice models can be formulated for both the measurement induced phase transition (MIPT) and the RTN \cite{vasseur2019entanglement,jian2020measurement,bao2020theory}. 
Each case has the $G_N$ layer-permutation symmetry that we met in Sec.~\ref{sec:entanglement:pairings}, but the required replica limits are ${N\rightarrow 1}$ for the MIPT and ${N\rightarrow 0}$ for the random tensor network.
This difference arises from the fact that  averages for the MIPT, unlike the RTN, must include the Born probability $\bra{\psi}K_{\bf m}^\dag K^{\phantom{\dag}}_{\bf m}\ket{\psi}$. This factor involves one layer each of the circuit $K_{\bf m}$ and its conjugate $K_{\bf m}^*$ (see Sec.~\ref{sec:entanglement:pairings}), so it increases $N$ by~1.

It is possible to view the models of interest, with $G_N$ symmetry, 
as symmetry-breaking perturbations of a  Potts model with $Q=N!$ states \cite{vasseur2019entanglement}.
This means initially treating all the $N!$ possible values for $\sigma$ as equivalent, giving a  larger $S_{N!}$ symmetry.\footnote{In the limits ${N\rightarrow 0,1}$ this becomes a $Q=1$ state Potts model \cite{vasseur2019entanglement}, which is a representation of percolation \cite{cardy1996scaling}. However --- confusingly --- this percolation fixed point is unrelated to the minimal cut limit discussed above (it does not correspond to a physical limit of the original problem).}
The  Landau-Ginsburg theory for Potts
is well-understood \cite{amit1976renormalization}.
However, reducing the symmetry to the physical symmetry,  ${G_N\subset S_{N!}}$, introduces an infinite number of relevant perturbations that are not related to each other by $G_N$ symmetry. This means that it is not easy to control the RG flow away from the Potts fixed point.

The basic issue is that $\sigma$
splits into an infinite number of distinct representations\footnote{Here we regard $\sigma$ as an element of the group algebra, i.e. we allow ourselves to take linear combinations as is natural if we want to coarse-grain.} of $G_N$, and one must decide  which should be retained as fundamental fields in a Lagrangian~\cite{nahum2021measurement}.
An alternative approach is motivated by the picture of overlaps between Feynman trajectories discussed in previous sections. We can introduce an Edwards-Anderson-like matrix order parameter, $X_{ab}$, which characterizes the strength of overlap between forward layer $a$ and backward layer $b$ \cite{nahum2021measurement}.
(Different values of $\sigma$ correspond to different ordered states for $X$.)
$X$ transforms simply under $G_N$, and using this one can write putative Landau-Ginsburg Lagrangians $\mathcal{L}(X)$ for the MIPT and RTN. It remains to be seen whether these conjectured theories describe the physical problems of interest.

Given these complications, it is natural to look for ways to simplify the transition. One is to get rid of spatial locality \cite{gullans2020dynamical,vijay2020measurement,nahum2021measurement}.
As noted above, there is no meaning to volume versus area law in all-to-all models, but purification (the amount of information  propagated between  initial and final times) can be used to distinguish the phases.

In the limit $q\rightarrow \infty$, 
 the all-to-all model mentioned at the end of Sec.~\ref{sec:purification}
maps to a min cut problem in a classical graph with temporal, but not spatial, locality. This is solvable essentially by percolation mean-field theory \cite{nahum2021measurement,gullans2020dynamical}.  A model with ``instantaneous quantum polynomial time'' gates \cite{bremner2011classical} shows another transition, described by mean-field percolation on a time-slice \cite{vijay2020measurement}.

At finite $q$, we can try to to exploit  the geometry of the quantum circuit.  
This is locally tree-like:  
the only loops have a size that diverges in the thermodynamic limit (as in many random graph ensembles \cite{bollobas1998random}).
This suggests that the MIPT   coincides with the entanglement transition of an  ensemble of \textit{tree tensor networks} \cite{lopez2020mean, nahum2021measurement} with the same local structure. 
If, as a simplification, the measurement outcomes  in the parent circuit are fixed using postselection 
(instead of sampled with Born's rule), 
this tree transition is exactly solvable  \cite{nahum2021measurement}.
The natural entanglement order parameter vanishes in a strongly non-mean-field fashion as ${\exp(-c/  \sqrt{p-p_c})}$. 
These all to all models are perhaps the simplest incarnations of the MIPT, and deserve further study.

\subsection{Structured monitored circuits}

There is a rich landscape of monitored dynamics beyond the minimally structured case of random two-qubit unitary gates interspersed with single site measurements.  { A simple extension is to consider dynamics with only multi-site measurements and no unitary gates.  Monitored pure-states in  measurement-only dynamics can also display phase transitions between volume and area law-scaling of the entanglement entropy \cite{ippoliti2021entanglement}.  This is  striking because there are no unitary gates in these models to ``compete'' with the measurements:
rather, the ``scrambling" and ``un-scrambling" are inextricably intertwined and the principle driving the transition is the mutual incompatibility or frustration of the measurement operators.}  

{ A further extension of unstructured, monitored dynamics is } to consider {symmetry enriched monitored dynamics}, obtained by restricting the unitary gates and measurements to operations that respect an on-site global  symmetry $G$ \cite{sang2021measurement,li2021robust,lavasani2021measurement, ippoliti2021entanglement}. Such models can still display an entanglement phase transition from area- to volume- law entangled steady state ensembles -- however, the symmetry structure of the replicated statistical mechanical description is enlarged by combining the circuit symmetry $G$ with the intrinsic dynamical symmetries of the problem (cf. Eq.~\ref{eq:replicasymm}). This permits novel types of dynamical orders that transcend the phase classifications obtained in the more conventional setting of static (or even Floquet) Hamiltonians with the same symmetry $G$. One upshot is that we can obtain multiple  varieties of both area and volume law phases -- for instance, distinguished by the presence or absence of different types of long-range symmetry-breaking orders or symmetry protected topological (SPT) orders --- with phase transitions between these occurring \emph{within} the area and volume law entangled phases.  Importantly, these orders are only observable in non-linear Edwards-Anderson type order parameters which measure fluctuations across trajectories, while simple averages remains featureless. 

It is illustrative to discuss an example of a phase transition between area-law states in a system with Ising symmetry $G = \mathbb{Z}_2$ \cite{sang2021measurement}. This example is reminiscent of a ground state phase transition from a paramagnet to a symmetry-broken ferromagnet, although a closer analogy may be the eigenstate phase transition between spin-glass ordered and paramagnetic states in many-body localized systems~\cite{huse_lpqo, pekker_hilbertglass}. We consider {measurement-only} Clifford dynamics in a one dimensional spin $1/2$ system. There are no unitary gates, and measurements are drawn randomly in space-time from two sets of commuting operators: with probability $p_Z$ from $\{Z_i Z_{i+1}\}$ and with probability $p_X = 1-p_Z$ from $\{X_i\}$. Each operator in the ensemble commutes with the symmetry generator $P=\prod_i X_i$, and we start with a symmetric initial state such as a product state in the $X$ basis. When $p_Z = 1$, the dynamics measures $Z_i Z_{i+1}$ on each bond, which projects onto Schrodinger ``cat states" with long-range spin-glass order and area law entanglement. A particular trajectory leads to a random sequence of measurement outcomes, $m_i = \pm 1$, on each bond; there are two symmetry broken product states in the $Z$ basis consistent with these outcomes (and related by the action of $P$), and the output state is a symmetric/antisymmetric cat superposition of these states, depending on the symmetry of the initial state. For example, if $m_i = \{+1,-1,+1\}$ in a system of length $L=4$, the output state is $|\psi_{\vec{m}}\rangle \propto |\uparrow \uparrow\downarrow\downarrow\rangle \pm |\downarrow \downarrow\uparrow\uparrow\rangle$. The ``glassy" order refers to the random orientation of spins (which generalizes the aligned pattern of a conventional ferromagnet); the long range order is diagnosed by a Edwards Anderson order parameter, $\chi^{\rm SG} = \lim_{L\rightarrow \infty} \sum_{\vec{m}} p_{\vec{m}} \frac{1}{L^2} \sum_{ij} \langle \psi_{\vec{m}}|Z_iZ_j| \psi_{\vec{m}} \rangle^2 >0$.
In contrast, when $p_X=1$, we measure $X_i$ on every site, and the output state is a paramagnetic product state in the $X$ basis with $\chi^{\rm SG} \rightarrow 0$. There is no volume-law phase in this model, and a dynamical phase transition between the paramagnetic and ordered area-law states occurs at $p_Z=0.5$. 

When the discrete Ising symmetry is replaced by a continuous symmetry, such as a global $U(1)$ symmetry, the phase structure appears to be even richer \cite{agrawal2021entanglement}.  For a $U(1)$ circuit with symmetry respecting two qubit unitary gates and single site $Z_i$ measurements, the volume law phase has been predicted to break into two phases, a ``charge-fuzzy" phase and a ``charge-sharp" phase.    In the former, which occurs at small measurement rate, $p$, an initial pure state which is a linear combination of different charge sectors (for example all spins pointing in the $X-$direction) will evolve into a charge eigenstate on times that are linear in the number of qubits.  In the charge-sharp phase
at higher $p$ (but still in the volume law phase) on the other hand, this charge sharpening occurs on times that are order one for large system size.  One can also consider an ancilla coupled to two different charge sectors, namely
$|\Psi \rangle = |\psi_Q \rangle |0\rangle + | \psi_{Q-1} \rangle |1\rangle$, with $|\psi_Q\rangle$ representing a state in the charge $Q$ sector (while $|0\rangle, |1\rangle$) are states of the ancilla).  In this case the ancilla qubit purifies on time scales of order one in the charge-sharp phase, but more slowly, of order $L$ in the charge-fuzzy phase.  { A proposed field theory predicts an infinite-order (Kosterlitz-Thouless) phase transition between the charge-sharp and charge-fuzzy phases \cite{barratt2021field}}.  Further explorations of circuits with $U(1)$ symmetry, or other non-Abelian symmetries, constitutes an exciting future direction. 

Monitored dynamics can drive phase transitions between area-law-entangled states with distinct topological quantum orders.  As an example \cite{lavasani2021topological, ippoliti2021entanglement},  measurements of the stabilizers of the toric code \cite{kitaev2003fault}, 
along with a weak rate of single-qubit measurements in the Pauli basis, can give rise to a topologically-ordered phase in which the monitored pure-states sustains long-range entanglement which cannot be removed by a finite-depth unitary circuit, as quantified by a topological entanglement entropy \cite{kitaev2006topological,levin2006detecting} of  $S_{\mathrm{topo}} = 2\ln 2$ on the torus.  Equivalently, in the purifying dynamics of a maximally-mixed initial state, the entanglement entropy of the system saturates to $S = S_{\mathrm{topo}}$ in constant time, and the system fails to completely purify up to times which scale exponentially in the linear dimension of the system. 

Monitored dynamics is also interesting for dynamics with free fermion structure \cite{cao2018entanglement,chan2019unitary,nahum2020entanglement,alberton2021entanglement,chen2020emergent,jian2020criticality,fidkowski2021dynamical,nahum2021measurement}, in which  unitary evolution and measurements of fermion bilinears only lead to the generation of two-body correlations. 
In contrast to a generic monitored dynamics,  free-fermion monitored evolution cannot sustain a volume-law-entangled phase for any non-zero monitoring rate in any number of spatial dimensions \cite{cao2018entanglement}. This is related to having a continuous rather than discrete replica symmetry, which reduces the cost of entanglement domain walls \cite{nahum2021measurement}. 
Equivalently, it has been shown that the purification time for an $N$-fermion state will be at most $O(N^{2})$, so that free-fermion monitored systems are always in a ``pure" phase \cite{fidkowski2021dynamical,nahum2021measurement}.

Apart from area-law-entangled steady-states, however, both the continuous- and discrete-time monitored dynamics of free fermions can give rise to super-area-law-entangled phases and critical points \cite{nahum2020entanglement,alberton2021entanglement,chen2020emergent}.

\subsubsection{Hybrid dynamics from spacetime-duality}
\label{sec:flipped_monitored}
We now consider a special class of monitored dynamics obtained via a spacetime rotation of unitary dynamics~\cite{ippoliti2021postselection, ippoliti2022fractal, lu2021spacetime}. These afford various benefits, both in the postselection cost of selecting quantum trajectories, and in furnishing a complementary analytic perspective relating monitored and unitary dynamics. 

As discussed in Sec.~\ref{sec:dualunitary}, viewing a unitary quantum gate $U_{i_1i_2}^{o_1o_2}$ sideways generically results in a non-unitary map, $\tilde{U}_{i_1o_1}^{i_2o_2}$. This map implements a \emph{forced} or postselected measurement: a specific unitary gate yields a specific fixed measurement outcome with no Born randomness. For example, if $U$ is a two-site identity gate, its dual $\tilde{\mathbb{1}} = 2\ket{B^+}\bra{B^+}$ is proportional to a projection onto a \emph{specific} Bell pair state $\ket{B^+} \equiv \frac{1}{\sqrt{2}} (\ket{\uparrow\uparrow} + \ket{\downarrow\downarrow})$\footnote{In general, a polar decomposition yields $\tilde{U} = 2 F W$, where $W$ is a unitary gate and $F$ is positive semi-definite and normalized to $\Tr(F^2) = 1$. 
Since $F\geq 0$, we can interpret $\tilde{U}$ as an element of a POVM: it corresponds to a \emph{forced weak measurement} (i.e., deterministically postselecting a particular outcome of a POVM).}. Performing this exchange in the roles of space and time across the entire circuit associates to every unitary evolution a non-unitary partner. The input and output states of the dual monitored evolution live on timelike slices and correspond to spatial boundary conditions of the unitary evolution, as shown in Fig. \ref{fig:dual_unitary}c.

Thus far, this seems to be a purely theoretical construction - after all, an experimentalist cannot directly implement forced measurements like $\tilde{\mathbb{1}}$ (except by costly postselection). However, a simple protocol described in \cite{ippoliti2022fractal} uses a ``teleportation" protocol to transfer the time-like input/output states to conventional space-like slices at the cost of introducing additional ancilla qubits and unitary SWAP gates; the system and ancilla qubits are then evolved with purely unitary gates (which the experimentalist \emph{does} have access to); following this evolution, a set of postselected Bell measurements at the final time-slice produces the desired output state corresponding to sideways hybrid evolution as shown in Fig. \ref{fig:dual_unitary}d. Thus, while these circuits do not eliminate the postselection problem, they parametrically improve the cost by only requiring postselection at the final time rather than the entire spacetime volume; this is also desirable for various near-term experimental architectures that do not allow measurements in the middle of the circuit but only at the end. 

Separately, flipped circuits provide a useful analytic perspective on monitored circuits by bootstrapping to the vast body of knowledge on temporal entanglement dynamics in unitary circuits. To zeroth order, spacetime duality exchanges the roles of space and time; hence, \emph{spatial} scaling of entanglement in the late-time states of flipped monitored circuits maps to the \emph{temporal} scaling of entanglement growth in the corresponding unitary circuit~\cite{ippoliti2022fractal, lu2021spacetime}. If the unitary circuit is chaotic and displays ballistic entanglement growth $S(t)\sim v_Et$, this translates to a volume law scaling for steady states in the flipped circuits, with an entropy density set by $v_E$, $\tilde{S}_A(\tilde{t}\rightarrow \infty) \sim v_E|A|$. Interestingly, this also implies that the variety of temporal entanglement dynamics in unitary settings (ranging from logarithmic to subballistic growth in time) translate to different spatial scalings in the output states (ranging from logarithmic to fractal) leading to new classes of entanglement phases in monitored dynamics. 

Importantly, however, the interchanging of space and time is not the full story.  The spatial scaling entanglement of subsystems of output states of monitored dynamics is mapped to the temporal growth of entanglement in unitary circuits, but the unitary evolution is accompanied by \emph{boundary decoherence}, which allows information to escape the system and be ``radiated away" from one of its edges~\cite{ippoliti2022fractal}. This furnishes a connection between monitored and dissipative dynamics, which were contrasted in Sec.~\ref{sec:trajectoryvsmixed}. The presence of boundary dissipation furnishes universal subleading corrections to the leading entanglement scaling which, for instance, characterize the non-thermal nature of the volume-law phase. For example, the entropy of a mixed state evolving under Haar random unitaries subject to boundary dissipation can be calculated using the domain-wall picture described in Section~\ref{sec:entanglement:pairings}, with the domain wall pinned to the boundary of the system at the final time. The edge dissipation changes the random walk calculation discussed earlier by introducing a partially absorbing boundary condition. This analysis furnishes subleading corrections to the temporal entanglement dynamics, coming from both the $t^{1/3}$ KPZ corrections present in the quenched average of entanglement, and an additional ${3/2}\log(t)$ piece from the partially absorbing boundary conditions. Upon dualizing, these give the same $|A|^{1/3}$ spatial corrections to the entanglement entropy of the volume law states obtained via the DPRE picture in Sec.~\ref{sec:measurementdwsec}; indeed, in this setting, one could think of the domain wall picture  as a microscopic realization of the (DPRE) effective description.

\section{Experiments} \label{sec:Experiments}

The subject of this review is  topical in light of rapid experimental progress in building programmable digital quantum simulators. The ability to isolate quantum coherent qubits,  to couple qubits via controlled unitary operations, and to make high-fidelity locally resolved measurements for readout, control and feedback all represent major engineering challenges; these challenges are being actively pursued over a wide range of physical platforms ranging from superconducting junctions to trapped ions. While much of this effort is broadly motivated by the quest to build universal programmable quantum computers --- a goal that is still far in the future given current parameters for noise rates and system sizes~\cite{preskill2018quantum} --- these platforms have already furnished impressive new capabilities when viewed as experimental platforms for many-body physics. This dual view of a computational device as an information theoretic tool on one hand, and a real experimental system for many-body physics on the other, is reflected in the broad interdisciplinary theoretical interest in quantum circuits. 

One of the first experimental breakthroughs in digital simulation was Google's announcement of ``quantum supremacy" in a 53 qubit system \cite{arute2019quantum}, signaling a leap from a decades long effort in designing and benchmarking individual quantum circuit elements to the arrival of genuinely many-body coupled systems with vast Hilbert spaces. An abstract information theoretic task was chosen for the demonstration, that of sampling from the output distribution of a state evolved under a random quantum circuit  \cite{neill2018blueprint,bouland2019complexity,aaronson2017complexity,boixo2018characterizing}; the task illustrates the  utility of quantum circuit dynamics for benchmarking near-term quantum devices.

From the point of view of many-body physics, it is particularly interesting to study  phenomena in the  regimes opened up by the natural operational mode of these devices that is accessible in the present-term i.e. viewing them as quantum circuits executing non-equilibrium dynamics~\cite{Ippoliti_nisq} (as opposed to universal simulators that may eventually shed light on long-standing equilibrium problems in strongly-correlated physics, such as the phase diagram of the 2D Hubbard model). These platforms are also building remarkable capabilities to access the new types of information theoretic observables we have discussed in this review. Tomographic techniques, while strongly limited to small system sizes, allow the full reconstruction of a density matrix - including quantum coherent off-diagonal terms - and thereby allow the computation of any non-linear function of the density matrix, including entanglement. A beautiful experiment on an (analog) Bose-Hubbard simulator \cite{islam2015measuring, Kaufman_2016} computed the purity, $\langle \mbox{Tr}\rho_A^2\rangle = \mbox{Tr}[ \mbox{SWAP}_A \rho \otimes \rho]$, by making two identical copies of a system and measuring a partial SWAP operator between the two copies \cite{daley2012measuring,alves2004multipartite}, circumventing the need for tomography at the expense of needing to prepare multiple copies.  Novel ancilla assisted measurement protocols can implement methods (like the ``Hadamard test") to obtain new types of correlation functions.

These capabilities have been put to use in several notable recent works. We focus here on experiments with quantum circuits, but note that these works follow many milestone papers probing fundamental aspects of quantum dynamics, thermalization and many-body localization on a wide variety of analog simulator platforms~\cite{Kaufman_2016, schreiber2015observation, Monroe_MBL, Bloch_MBL_2D, Greiner_entanglement_mbl, Bernien_2017, Weiss_newtoncradle, Lev_integrability}. Ref.~\cite{Google_scrambling} furnished a detailed experimental study of information scrambling in a variety of chaotic quantum circuits by measuring out-of-time-ordered commutators using ancilla-assisted methods. Ref.~\cite{Google_toric} implemented a circuit to prepare the ground state of the topologically ordered toric code and performed a measurement of the topological entanglement entropy of the state and simulated an anyon braiding operation on the state. Ref.~\cite{Google_TC_Exp} implemented an MBL Floquet circuit to simulate a time-crystal, and made an ancilla assisted measurement of a {spectrum averaged} unequal space-time correlation function closely related to the Edwards-Anderson correlator discussed previously; this correlator probes the defining spatiotemporal order of the phase across the \emph{entire} exponentially dense many-body spectrum, and contrasts with conventional correlators that vanish on taking an ``infinite temperature" averaged over the entire spectrum. 

When considering monitored circuits with unitaries and measurements, we must reckon with the prohibitive postselection barrier mentioned earlier. Here, the challenge of preparing multiple copies of a given quantum state (i.e. a given trajectory associated with a specific set of measurement outcomes $\{\bf m\}$) is not ``merely" an engineering one, but a fundamental theoretical one stemming from the randomness inherent to the measurement process. Ref.~\cite{noel2021observation} made a first experimental attempt to probe this physics in a small 8 qubit trapped ion system by measuring the late-time entanglement entropy of a ``reference qubit" entangled with the system at $t=0$, as discussed in Section.~\ref{sec:purification}. This method requires the existence of a ``decoder" to correlate the basis in which the reference qubit is measured with the measurement record on the system, which was achieving in the experiment by simulating Clifford circuits whose classical simulability allows ``wrong" measurement outcomes to be corrected by a feed-forward action determining future unitary operations. On small enough systems, individual trajectories can be reproduced by ``brute-force", as was done in a recent experiment using IBM's digital simulators \cite{koh2022experimental}.  
However, larger scale experimental demonstrations of non-Clifford monitored evolutions will require fundamental new approaches to address the postselection problem, for instance by appealing to additional structures like space-time duality that can parametrically reduce postselection overhead as discussed in Section~\ref{sec:flipped_monitored}.

\section{Outlook} \label{sec:Outlook}
We conclude by outlining some important topic areas and questions within quantum circuit dynamics that deserve further study.  First, the effects of \emph{feedback} in quantum dynamics -- whereby future unitary operations or measurements are conditioned on past measurement outcomes -- remain to be understood.  How can feedback be harnessed to stabilize ordered phases of quantum matter in quantum circuit dynamics, and how can these protocols be realized in digital quantum simulators?  How does feedback affect the generation of quantum many-body entanglement?  Quantum error-correction \cite{nielson2000quantum} provides one  well-studied example of the powerful effects of feedback, which can allow an observer to recover an unknown quantum state encoded in an evolving quantum system,  though a more extensive exploration of this new area of ``interactive" quantum dynamics is warranted.\footnote{We note that successful quantum error-correction would lead to the recovery of a  particular encoded quantum state with high fidelity, which is \emph{a priori} a much more stringent requirement than using feedback to stabilize a phase of a quantum many-body system.}   The out-of-equilibrium phases that can arise due to the interplay of monitored dynamics with open quantum system dynamics, which evolve the system of interest into a mixed state, is beginning to be explored \cite{li2021entanglement,weinstein2022measurement,noh2020efficient,li2022entanglement} and also provides a fruitful area of study.  

In the study of monitored quantum circuits, probing the entanglement properties of monitored pure-states generally requires a large amount of post-selection, as explained in Sec. \ref{sec:trajectoryvsmixed}, which poses a barrier for experimentally-observing the measurement-driven entanglement phase transition.  Are there classes of monitored dynamics, apart from Clifford dynamics and evolution with a dual-unitary description \cite{ippoliti2021postselection}, in which this ``post-selection barrier" can be avoided? Can spacetime rotations of unitary quantum circuits \cite{lu2021spacetime,ippoliti2022fractal} be used to overcome post-selection problems in the preparation of other interesting kinds of quantum states?  

The  entanglement phase transition in monitored dynamics remains to be fully characterized. In the Clifford case, a quantitative understanding of the evolution of stabilizers (Sec.~\ref{sec:clifford}) in monitored Clifford circuits might shed light on this transition.
Progress in understanding the relevant statistical mechanical descriptions of monitored systems with continuous symmetries would also provide useful insight into phases that can arise in this setting. 
As we touched on in Sec.~\ref{sec:monitored}, there are also interesting questions about coarse-graining the effective models for the case of generic unitaries,
with connections to fundamental concepts in the theory of disordered systems. 
Separately, it would also be  interesting to have mathematically rigorous results for the phase diagram of the measurement problem, for example a rigorous proof of the stability of the volume-law phase.

At the level of formalism there is also much still to understand about the structure of the effective lattice models for both unitary and nonunitary dynamics, for example about the combinatorial structure underlying various replica-like limits.
While we have discussed here a limited number of quantities, this formalism can be applied to almost any observable of interest in the circuit.

As we discussed at the outset, one way to motivate the circuits is as simpler models for dynamics in more conventional many-body systems with a fixed Hamiltonian, either on the lattice or in the continuum. In Sec.~\ref{sec:entanglement:pairings} and Sec.~\ref{sec:floquet}
we touched on some ways of extending ideas from the random circuit to non-random systems but there is much still to do here.

Finally, can monitored dynamics provide ($i$) new benchmarking tasks for near-term quantum computers or ($ii$) insights into quantum error-correcting codes?  Time-periodic monitored dynamics have been recently used to construct dynamically-evolving, fault-tolerant quantum codes \cite{hastings2021dynamically,aasen2022adiabatic}, though it remains to be understood if other kinds of monitored evolution can produce similarly useful codes, in which ``decoding" quantum information is practical and feasible.

\section*{DISCLOSURE STATEMENT}
If the authors have noting to disclose, the following statement will be used: The authors are not aware of any affiliations, memberships, funding, or financial holdings that
might be perceived as affecting the objectivity of this review. 

\section*{ACKNOWLEDGMENTS}
We thank Ehud Altman, Denis Bernard, Bruno Bertini, John Chalker, Amos Chan, Xiao Chen, Andrea De Luca, Fabian Essler, Ruihua Fan, Zongping Gong, Sarang Gopalakrishnan, Michael Gullans, Jeongwan Haah, Timothy Hsieh, David Huse, Matteo Ippoliti, Cheryne Jonay, Yaodong Li, Andrew Lucas, Andreas Ludwig, Roderich Moessner,
Rahul Nandkishore,
Lorenzo Piroli,
Tomaz Prosen, Tibor Rakovszky, Sthitadhi Roy, Jonathan Ruhman, Shengqi Sang, Brian Skinner, Shivaji Sondhi, Tobias Swann, Ashvin Vishwanath, Curt von Keyserlingk, Romain Vasseur, Zhi-Cheng Yang, Yi-Zhuang You, Tianci Zhou,
Marko Znidaric,
for collaboration on related work and/or fruitful discussions.

V.K. acknowledges support from the Sloan Foundation through a Sloan Research Fellowship, the Packard Foundation through a Packard Fellowship, and by the US Department of Energy, Office of Science, Basic Energy Sciences, under Early Career Award No. DE-SC0021111. V.K. thanks the Kavli Institute of Theoretical Physics (KITP) for hospitality during a part of this project. KITP is supported in part by the National Science Foundation under Grant No. NSF PHY-1748958.   M.P.A.F. gratefully acknowledges support from the Heising-Simons Foundation, and by the Simons Collaboration on Ultra- Quantum Matter, which is a grant from the Simons Foundation (651440).

\bibliographystyle{unsrt11}

\bibliography{references}

\end{document}